\documentclass[11pt,twoside]{article}
\usepackage{lscape}
\usepackage{latexsym}
\usepackage[lofdepth,lotdepth]{subfig}
\usepackage{graphicx}
\usepackage{amsmath}
\usepackage{amssymb}
\usepackage{amsthm}
\usepackage{epsfig}
\usepackage{bbm}
\usepackage[numbers,square]{natbib}
\usepackage{lgrind}
\usepackage{fullpage, amsfonts, longtable, algorithm, algorithmic}
\usepackage{graphicx,color}

\allowdisplaybreaks[1]
\usepackage{setspace}

\usepackage{float}
\usepackage{amssymb}
\usepackage{multirow}
\usepackage{array}
\usepackage{natbib, url}

\newtheorem{theorem}{Theorem}

\title{A New Framework for Network Disruption}

\author{S. E. Martonosi, D. S. Altner, M. Ernst, E. Ferme, K. Langsjoen, D. Lindsay, S. Plott, A. S. Ronan}

\date{\today}

\begin{document}
\maketitle

\begin{abstract}
Traditional network disruption approaches focus on disconnecting or lengthening paths in the network.  We present a new framework for network disruption that attempts to reroute flow through critical vertices via vertex deletion, under the assumption that this will render those vertices vulnerable to future attacks.  We define the \textit{load} on a critical vertex to be the number of paths in the network that \textit{must} flow through the vertex.  We present graph-theoretic and computational techniques to maximize this load, firstly by removing either a single vertex from the network, secondly by removing a subset of vertices.
\end{abstract}

\section{Introduction}

Network disruption has important applications to telecommunications, energy transmission, robust network design and counterterrorism.  While most studies on network disruption examine the effects of vertex deletion on network connectivity and path lengths, very little work has examined the rerouting of flow through a network.  This motivates the following combinatorial optimization problem:  Consider a connected \textit{graph} $G=(V,E)$ with vertices $V$, and edges $E$. An edge is denoted by a pair of vertices $(u,v)$ and represents opportunities for flow between vertices $u$ and $v$.  For instance, if $u$ and $v$ represent cities, then an edge could be a road connecting the two; if $u$ and $v$ represent people, then an edge could represent a direct communication line between them.  Assuming the volume of flow passing through a vertex is a proxy for the vulnerability of that vertex, which vertices should we eliminate from the network to maximize a critical vertex's vulnerability?

In this paper, we present a new network diversion problem applied to social networks, prove some properties of this disruption approach, and illustrate its potential value through simulations and meta-heuristics.  Solving this problem on large-scale networks poses interesting computational challenges, such as creating heuristics for identifying suitable targets for removal from the graph. Additionally, this work could be used for disrupting covert networks, such as terrorist, weapons smuggling, illegal drug or human trafficking networks, as well as for designing robust communication networks.  Although only a model of true communication interactions, we expect this work to provide important insights into network disruption strategies.

In Section~\ref{sec:lit}, we frame this work against the context of the existing literature. Then, in Section~\ref{sec:framework}, we present our modeling framework with key definitions and assumptions.   Section~\ref{sec:single} provides theoretical and empirical results of removing a single vertex to increase the load on a critical vertex, and Section~\ref{sec:multiple} provides the results of using a genetic algorithm to choose larger subsets of vertices to remove.  Section~\ref{sec:apps} discusses one possible application of this work to counterterrorism, Section~\ref{sec:future} provides future extensions of this work, and Section~\ref{sec:conc} concludes.

\section{Literature review}
\label{sec:lit}

In this section, we contrast the network interdiction and diversion approaches used to date with those we will present in this paper. The two main distinctions will lead our discourse towards the relevant foundational work from social network analysis.

Network interdiction models address the logistical problem of inhibiting the flow of resources through a network, which has applications to military operations and combating drug trafficking. Analysis of complex network interdiction typically focuses on disconnecting the network, increasing the lengths of shortest paths, or cutting overall flow capacity in the network \cite{AJB, FFV, GABCH, GCABH, GCLABH, GMR, GMMS, HKYH, MHH, PSS, SLCY, WDTZ}.  The most well-known model involves maximum flow network interdiction \cite{AEU2010, CMW1998, MM1970, P1993, RoW, RSL1975, W1993}. A related problem to network interdiction is network diversion, where arcs are removed from the network so that all pairwise flow must be routed through at least one member of a pre-specified set of ``diversion'' arcs \cite{CCDEGOQ, Cur}.

There are two gaps in the existing network disruption and interdiction literature that our model attempts to address: the incorporation of network structure, and a focus on flow through vertices.

First, previous work does not focus on the structure of the graph being interdicted.  As a result, these models do not yield structural insights into which types of vertices or edges should be targeted.  Complex networks are large networks whose structure arises due to a random evolutionary process dictating how vertices (representing people, objects or ideas, for instance) become interconnected over time.  Complex networks modeling human interactions are called \textit{social networks}.  Network evolution models have been developed to simulate networks observed in our world (e.g. the Internet, or social networking websites).

The following four common types of network models are used in our research:

\noindent \textit{\textbf{Erd\H{o}s-R\'{e}nyi random graphs:}}
This simplest random graph model begins with a collection of vertices and creates an edge between each pair of vertices independently with probability $p$ \cite{ErR}.  This random graph model bears little resemblance to natural or man-made complex networks, but is a useful benchmark for comparisons with other models.

\noindent \textit{\textbf{Watts-Strogatz small world graphs:}}
Many large social networks exhibit the ``small world'' property: path lengths between randomly selected pairs of vertices tend to be small \cite{New:2000, New:2003, NMW, NSW, NeW:1999PhysLet, NeW:1999PhysRev, NWS, Wat:1999, Wat:SmallWorld, Wat:SixDegrees, Wat:2007, WaS}.  The Watts-Strogatz model for generating such graphs begins with a ``ring'' graph in which each vertex is connected to its $k$ nearest neighbors to the left and to the right.  Next, each edge is randomly rewired with probability $p$.

\noindent \textit{\textbf{Barab\'{a}si-Albert power law graphs:}}
Many large-scale complex networks exhibit a power law asymptotic degree distribution, in which there are a few ``hub'' vertices with very large degree.  Preferential attachment is often used to explain the evolution of such graphs \cite{BaA, BAJ} (though other models can also be constructed; e.g. see \cite{DALLRSTW, LADW}). In this model, a new vertex $v$ links to an existing vertex $w$ with probability proportional to the degree of $w$.  The parameters are $m_0$, the number of vertices used to initialize graph generation, and $m$, the number of edges generated by each new vertex added to the graph.

\noindent \textit{\textbf{Holme-Kim power law graphs with clustering:}}
The Barab\'{a}si-Albert model does not exhibit the clustering common to many social networks.  Clustering arises when $A$ being connected to both $B$ and $C$ increases the likelihood that $B$ and $C$ will also be connected. Holme and Kim \cite{HoK} added a clustering step to Barab\'{a}si and Albert's preferential attachment algorithm to increase the prevalence of clusters while maintaining an asymptotic power law degree distribution.

    The importance of certain vertices in a social network is measured with \textit{centrality} metrics.  There are many such metrics (see, for example,  \cite{WaF}), but the most commonly used metrics are \textit{degree, betweenness} and \textit{closeness}.  The degree of a vertex is the number of neighbors it has.  The betweenness of a vertex is the number of shortest paths between all pairs of vertices on which the vertex lies.  Closeness measures the average shortest path length between the vertex and all other vertices in the graph.

The second gap of existing research is that the objectives of disconnecting the graph or increasing shortest path lengths are not always appropriate.  For instance, disconnecting a power network might require more interdiction resources than are available, but rerouting excessive power through a critical transmission vertex could result in a failure throughout the network. Covert networks tend to communicate along longer paths that are difficult to trace,  suggesting a trade-off between efficiency and secrecy \cite{FBW, MGP} that could render path-length-based attacks ineffective.  Thus, we define a new framework for network disruption based on network flow that could be applicable in contexts left unaddressed by the current literature.  We discuss this framework in the next section.

\section{Modeling framework} \label{sec:framework}

We can take advantage of the structural characteristics of social networks to obtain stronger network disruption strategies.  To our knowledge, no previous work has applied network diversion to social networks.  Moreover, rather than use \textit{length} as a success metric, we use the fact that critical vertices in certain types of networks can become vulnerable if they engage in greater \textit{quantities} of activity.  Thus, we identify vertices to remove from the network in such a way as to force more flow through a critical vertex.  This is the novel contribution of this research.

We start by presenting terminology that will be used in the paper.

\subsection{Preliminaries}

Given a graph $G = (V, E)$, we assume the edges in $E$ are \textit{undirected}.  We say that two vertices are \textit{adjacent} if they share an edge, and we say that an edge is \textit{incident} to a vertex if that vertex is one of the endpoints of the edge.

A \textit{$u$-$v$ path} is a sequence of edges $e_1, e_2, \ldots, e_m$ that connects vertex $u$ to vertex $v$ while passing through every other vertex at most once. More specifically, $e_1$ is incident to $u$, $e_m$ is incident to $v$, edges $e_i$ and $e_{i+1}$ are incident to a common vertex for each $i$ in $\lbrace 1, 2, \ldots m \rbrace$ and each vertex is traversed at most once when the $u$-$v$ path is traversed. A set of paths is called \textit{edge-disjoint} if the intersection of their edge sets is empty.

If we assume that each edge has a maximum capacity of flow that can pass along it, then we can define the \textit{maximum flow between vertices $u$ and $v$} to be the largest amount of flow that can travel from $u$ to $v$ along paths in the network such that the total flow along any edge does not exceed the maximum capacity of that edge.  In this paper, all edges will have unit capacity, so a maximum flow between $u$ and $v$ will be equivalent to a maximal set of edge-disjoint $u$-$v$ paths.

\subsection{Network flow centrality: ``load''}
As described earlier, there are several metrics for \textit{centrality}, or the importance of a vertex in a network.  We focus in this paper on \textit{network flow centrality}~\cite{FBW}, a special case of which we will call \textit{load}, for short, and define later in this section.

Each graph has a \textit{key vertex}, $k$, which could represent, for example, an important leader of an organization, or an important transmission junction in a power network. The objective is to identify a set of vertices to remove from the graph to make the key vertex $k$ as ``active'' as possible by forcing flow to pass through that vertex.

To measure the activity of the key vertex, we quantify how
much flow \textit{must} pass through it. A vertex is less critical for flow if there are many detours that avoid it. To count detours, we count edge-disjoint paths. Let $z_{st}(G)$ be the number of edge-disjoint $s$-$t$ paths in graph $G$. The \textit{flow capacity of graph $G$ with respect to key vertex $k$} is defined as

\begin{equation} \label{eqn:commcap}
Z_k(G) = \sum_{\substack{s,t \in V \setminus \{k \} \\ s \neq t}} z_{st}(G),
\end{equation}

\noindent and it is a metric for the total amount of flow that can be transmitted in graph $G$ that does not originate or end at $k$.

To ascertain how important an individual vertex $k$ is to the flow capacity of a graph, we introduce a metric called the \textit{load}, which equals the flow capacity of the graph with respect to $k$ minus the flow capacity with respect to $k$ of the subgraph obtained when vertex $k$ is deleted. More formally, the load of a vertex $k$ in a graph $G$ can be expressed as

\begin{equation} \label{eqn:load}
\mathcal{L}_k(G) = Z_k(G)-Z_k(G \setminus \{ k \}).
\end{equation}

\noindent $\mathcal{L}_k(G)$ counts the number of edge-disjoint $s$-$t$ paths that must include $k$ for all pairs of vertices $s$ and $t$ in $V \setminus \{k\}$.  Freeman, \textit{et al.}  \cite{FBW} define this as \textit{network flow
centrality} for the case of graphs with arbitrary
edge capacities.

To measure how the removal of a subset of vertices impacts the load of the key vertex, we define the \textit{load effect} of subset $S$ on key vertex $k$ to be the change in the key vertex $k$'s load caused by removing subset $S$, which can be formally stated as  \begin{equation} \label{eqn:loadeffect}
\mathcal{E}_k(G,S)=\mathcal{L}_k(G\setminus S) - \mathcal{L}_k(G ).
\end{equation}If the load effect of $S$ on $k$ is positive, then removing subset $S$ from the graph has diverted more flow through $k$, a desired effect.

\subsection{Load maximization problems}

The goal of this research is to identify the subset of vertices $S$ having the greatest load effect on a given key vertex $k$.  This is equivalent to choosing the subset $S$ which maximizes the value of $\mathcal{L}_k(G \setminus S)$.  We formally define the \textit{Load Maximization Problem} (LOMAX) as \begin{equation} \label{eqn:LOMAX}
max_{ S \subseteq V} \ \mathcal{L}_k(G \setminus S).
\end{equation}  We also study a special case of LOMAX where only a single vertex can be deleted, which we call the \textit{Single Vertex Deletion Load Maximization Problem} (Single-LOMAX).

LOMAX is a very difficult problem. Unlike network interdiction and network diversion problems, LOMAX cannot be formulated as an integer linear program since the load of the key vertex cannot be modeled with a linear function.  Indeed, there is no known functional form for the load effect of a subset on the key vertex, and load effect is not monotonic, or even convex, over subsets. Therefore, we must explore other possibilities for solving this problem.

We begin by studying the Single-LOMAX problem of identifying a single vertex having the greatest load effect on a key vertex $k$.

\section{Single vertex deletion (Single-LOMAX)} \label{sec:single}
From expressions (\ref{eqn:commcap}), (\ref{eqn:load}) and (\ref{eqn:loadeffect}) of the previous section, we see that there is no guarantee that the load effect on $k$ of removing vertex $i$, $\mathcal{E}_k(G, i)$, need ever be positive.  When vertex $i$ is removed from the graph, the overall flow capacity $Z_k(G\setminus \{i\})$ necessarily decreases because $i$'s contribution to the flow is removed.  In order for $i$'s removal to have a positive load effect on key vertex $k$, the remaining flow must be rerouted through $k$ in sufficiently large quantities to overcome the overall decrease in flow.

\begin{figure}[h!]
	\centering
	\qquad
	\subfloat[][Graph $G$]{
	\includegraphics[width=3in]{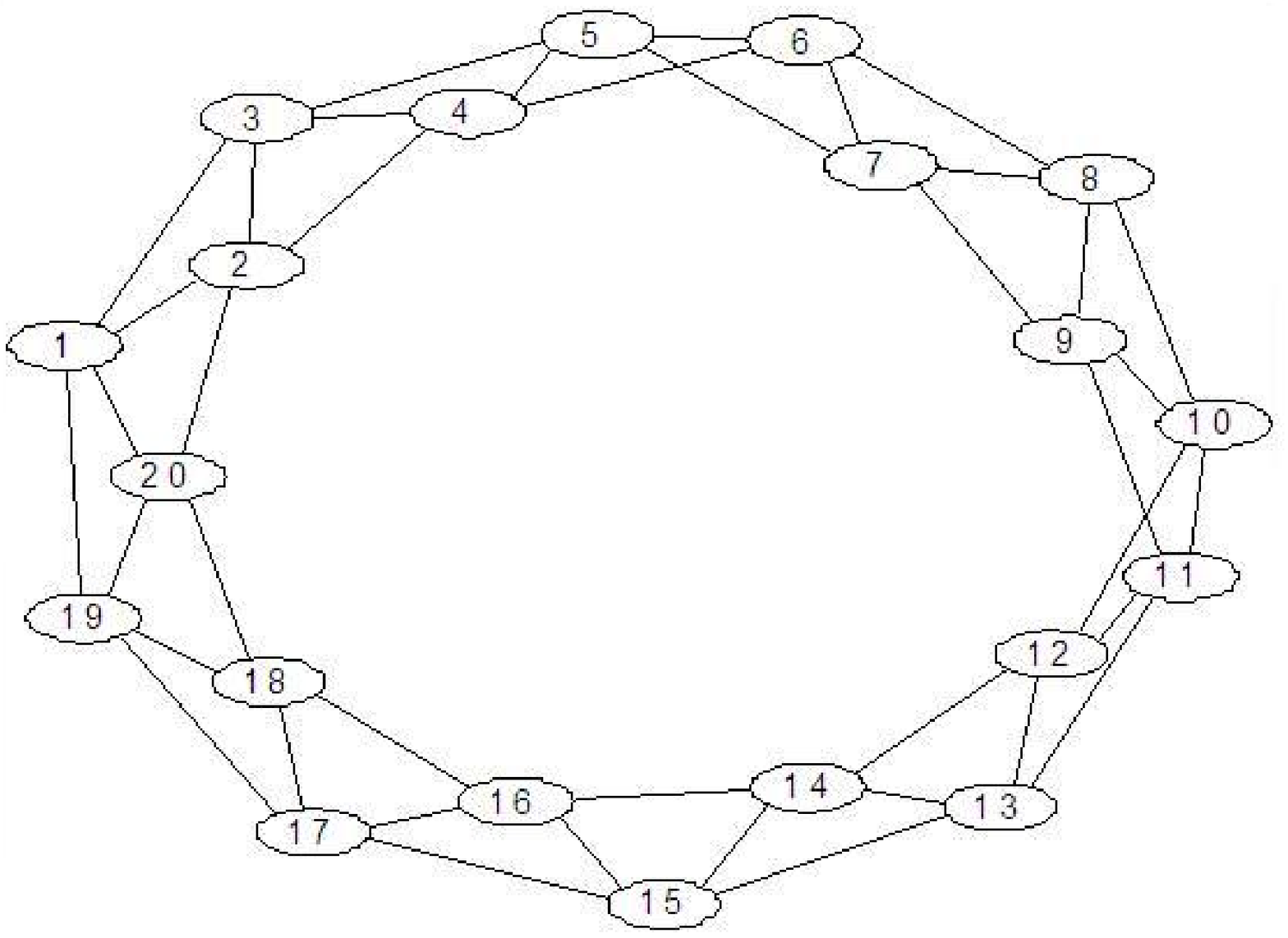}
	\label{fig:ProofConcepta}}
	\subfloat[][Graph $G \setminus \{k\}$]{
	\includegraphics[width=3in]{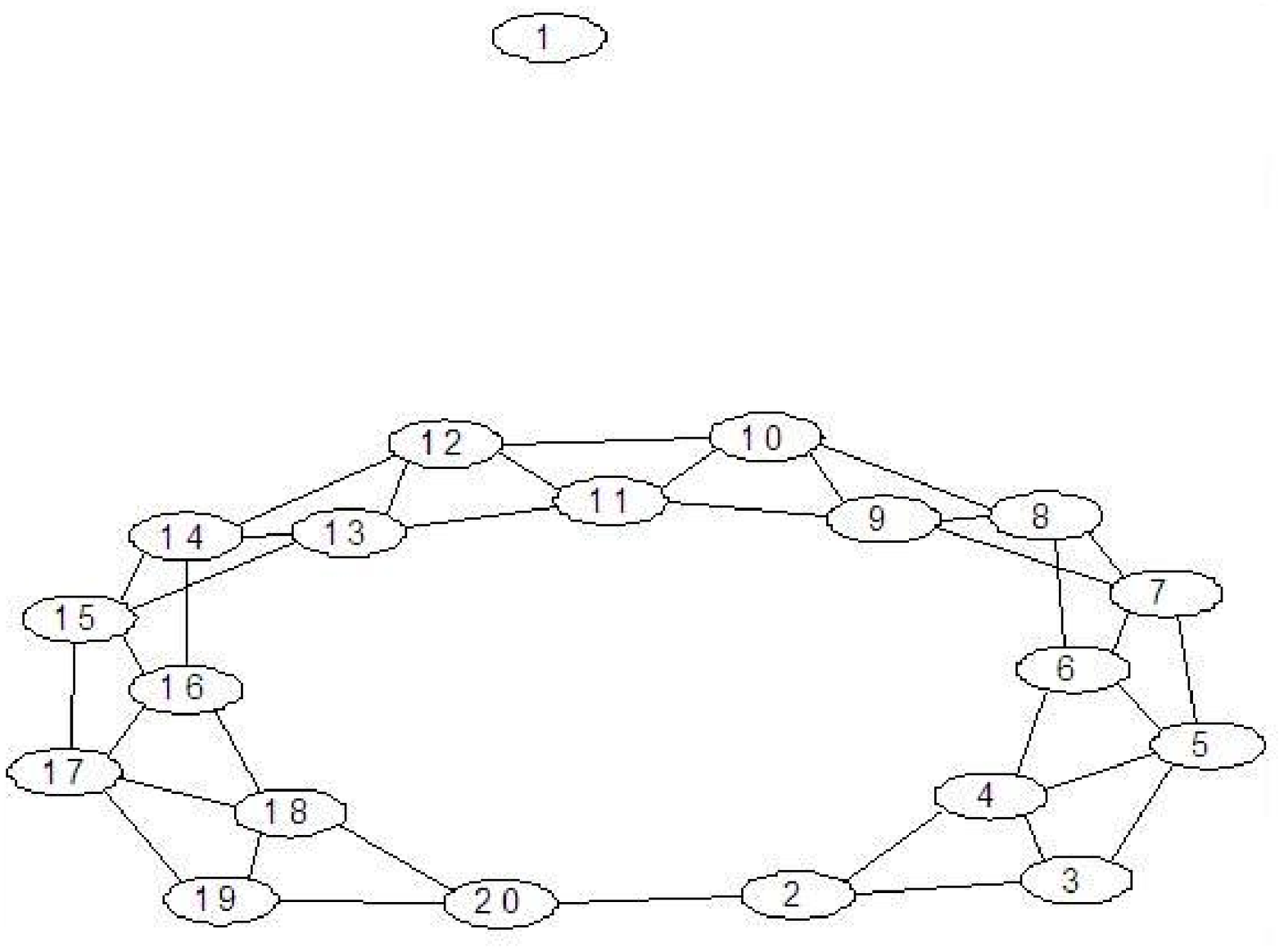}
	\label{fig:ProofConceptb}}
	\qquad
	\subfloat[][Graph $G \setminus \{i\}$]{
	\includegraphics[width=3in]{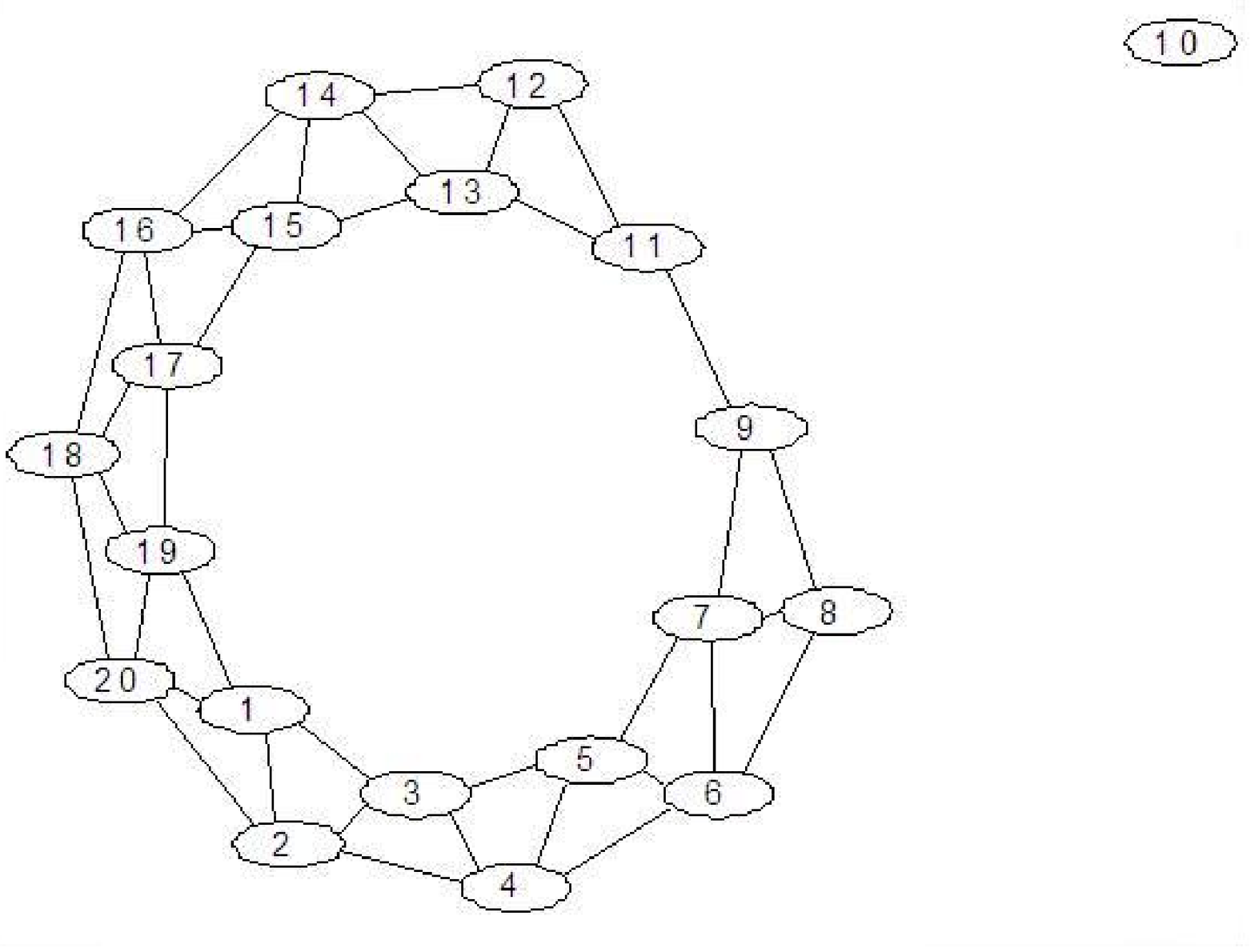}
	\label{fig:ProofConceptc}}
	\subfloat[][Graph $G \setminus \{i,k\}$]{
	\includegraphics[width=3in]{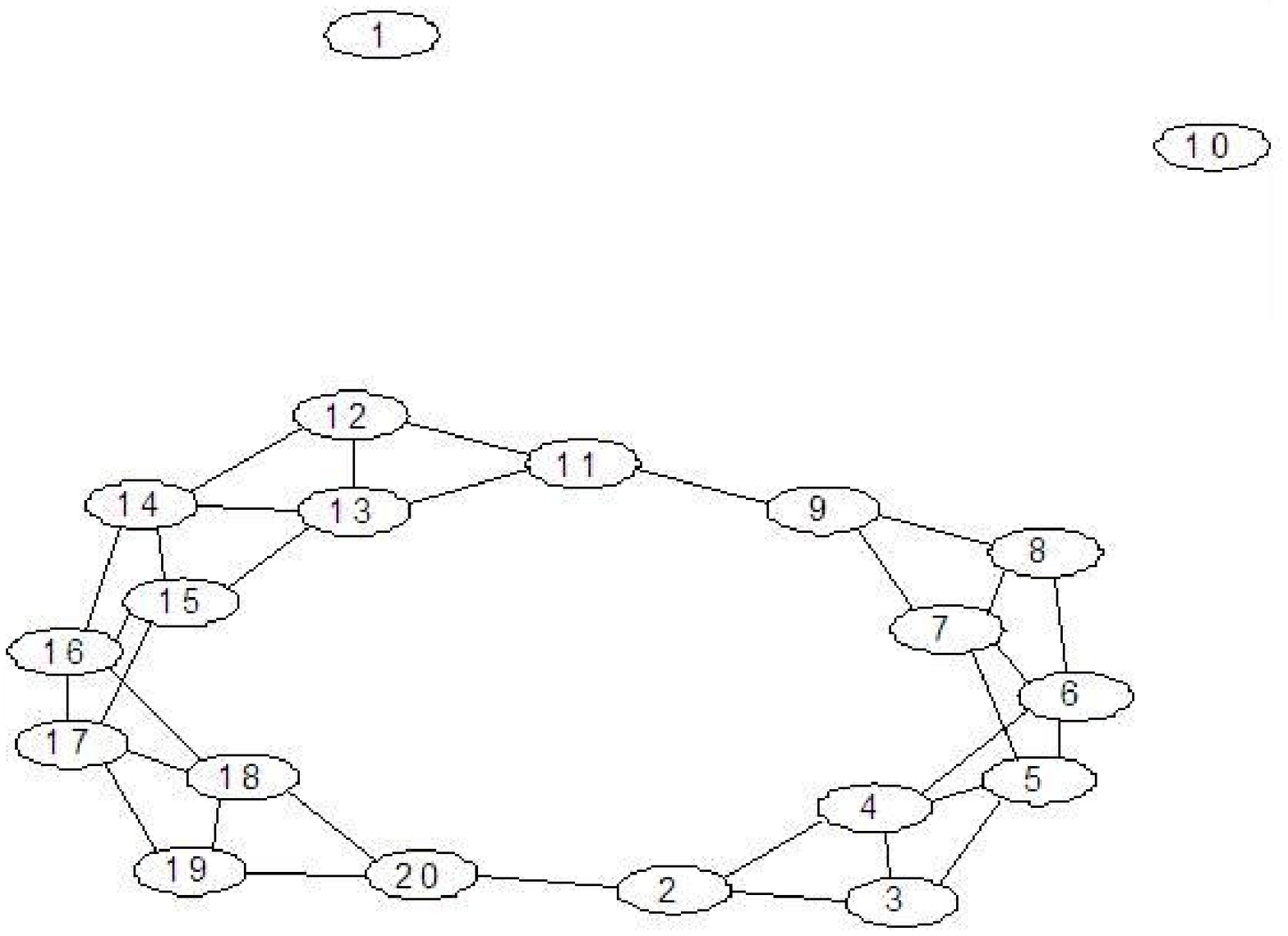}
	\label{fig:ProofConceptd}}
	\caption{Proof of concept demonstrating that removing a vertex ($i=10$) can increase the load on a key vertex ($k=1$). Figure \subref{fig:ProofConcepta} shows the original graph, $G$, with key vertex $k=1$.  This graph has a flow capacity with respect to $k$ of 684 paths.  Figure \subref{fig:ProofConceptb} shows graph $G \setminus \{k\}$, with key vertex $k=1$ removed.  Removing $k$ has reduced the flow capacity by 66 to 618 paths, so the load on $k=1$ in $G$ is 66. Figure \subref{fig:ProofConceptc} shows graph $G \setminus \{i\}$, with vertex $i=10$ removed.  The flow capacity in this graph is 550.  Figure \subref{fig:ProofConceptd} shows graph $G \setminus \{i,k\}$, with both the key vertex $k=1$ and vertex $i=10$ removed.  Removing $k$ has reduced the flow capacity by 150 to 400, so the load on $k$ in $G \setminus \{i\}$ is 150, which is 84 units higher than in the graph $G$.  Therefore, the load effect on $k$ of removing vertex $i$ is 84.}
	\label{fig:ProofConcept}
	\end{figure}

Figure \ref{fig:ProofConcept} gives an example demonstrating  it is possible to increase the load on a key vertex by deleting another vertex from the graph.  Figure \ref{fig:ProofConcepta} shows the original graph $G$, in which vertex 1 is the key vertex, $k$.  Figure \ref{fig:ProofConceptb} shows the graph $G \setminus \{k\}$.  Because 66 edge disjoint paths disappear when $k=1$ is removed from $G$, the load on $k$ in graph $G$ is 66.  Figures \ref{fig:ProofConceptc} and \ref{fig:ProofConceptd} perform the same calculations when vertex $i=10$ has been removed, and show that the load on the key vertex is now 150.  The load effect on the key vertex $k=1$ of removing vertex $i=10$ is therefore 84.

One question is whether a vertex having a positive load effect on the key vertex can be found in general.  If such vertices can generally be found, another question is how to identify such vertices.  In Section \ref{subsec:bruteforce} we provide evidence through brute force search that suggests that a vertex whose removal has a positive load effect on the key vertex can almost always be found when the key vertex is itself highly central.  However, characterizing such vertices is difficult, as we discuss in Section \ref{subsec:difficulties}.  We offer theoretical results that characterize when a vertex is guaranteed not to have a positive load effect in Section \ref{subsec:proofs}, and present some heuristic approaches in Section \ref{subsec:heuristics}.

\subsection{Proof of concept} \label{subsec:bruteforce}
We have empirical results demonstrating that nearly all graphs have at least one vertex whose load effect on the key vertex is positive. Specifically, we randomly generated instances from the four classes of social network graphs mentioned in Section~\ref{sec:lit}. For each graph, we selected the key vertex $k$ to be the vertex that had the highest average rank over three centrality types: betweenness, closeness and degree centrality. We then solved Single-LOMAX by brute force to identify the vertex whose removal had the greatest load effect on the key vertex.

For every graph type, we generated 276 instances of 100-vertex graphs.  The Erd\H{o}s-R\'{e}nyi random graphs used a rewiring probability of $p = 0.10$.  This resulted in a density of edges of 10\%, an average shortest path distance of 2.2, and a clustering coefficient of 0.10.  The Watts-Strogatz small world graphs were generated from ring graphs in which each vertex was connected to its $k=2$ nearest neighbors, and then edges were randomly rewired with probability $p=0.10$.  This yielded graphs with density 4\%, average shortest path length of 5.1 and a clustering coefficient of 0.38. The Barab\'{a}si-Albert power law graphs were generated using $m_0=3$ initial vertices and $m=2$ edges generated by each new vertex added to the graph.  They had an edge density of 4\%, an average shortest path length of 3.0 and a clustering coefficient of 0.12.   The Holme-Kim power law graphs with clustering were generated using $m_0=2$ initial vertices and $m=2$ edges generated by each new vertex added to the graph.  They had an edge density of 10\%, an average shortest path length of 2.3 and a clustering coefficient of 0.38.

\begin{table}
\begin{center}
{\scriptsize \begin{tabular}{|b{2.5cm}|| c c c c || c c  c c||c c  c c||}
\hline
Graph & \multicolumn{4}{c||}{Original Load on Key Vertex} & \multicolumn{4}{c||}{Average \% Load Effect (\%)}  &  \multicolumn{4}{c||}{Best \% Load Effect (\%)} \\
Type & Min & Med. & Mean & Max  & Min & Med. & Mean & Max & Min & Med. & Mean & Max\\
\hline
Erd\H{o}s-R\'{e}nyi &&&&&&&&&&&& \\
random $(p=0.1)$  & 415 & 814.5 & 822.5 & 1300& -12.2 & -1.4 & -2.0 & 3.4 & 0.8 & 3.2  & 3.4  & 8.9  \\
\hline
 Watts-Strogatz small world &&&&&&&&&&&& \\
  $(p=0.1, k=2)$ & 62 & 421.5 & 503.9 & 3168 & -35.8 & -3.8 & 10.7 & 180.8 &  17.3& 130.5 & 180.8 & 1701.4 \\
 \hline
Barab\'{a}si-Albert power law &&&&&&&&&&&& \\
$(m_0=3, m=2)$   & 507 & 1278.5 & 1301.7 & 2425 & -12.4& -1.3&-1.7 & 19.3 & 1.3 & 16.4 & 19.3  & 68.1 \\
 \hline
Holme-Kim power law w.~clustering $(m_0=2, m=2)$ & 903 & 1537.5  & 1548.7  & 2260  & -6.5&  -1.6&-2.0 & 1.2 & 0.05 & 1.0 & 1.2 & 5.3  \\
\hline
\end{tabular}
\caption{Load effect on a key vertex by graph type. For each graph type, we give the minimum, median, mean and maximum of the following fields across 276 graphs: the original load on the key vertex, the average load effect expressed as a percentage relative to the key vertex's original load, and the maximum load effect of a single vertex expressed as a percentage relative to the key vertex's original load. \label{table:removal}}}
\end{center}\end{table}

The results of these computations are given in Table \ref{table:removal}.   We henceforth define the average load effect to be the arithmetic mean of the load effect of each vertex in a single graph on that graph's key vertex.  The `Average \% Load Effect' is then the ratio of a graph's average load effect over the original load on that graph's key vertex. As seen in the `Mean' column for `Average \% Load Effect' in Table \ref{table:removal}, the average load effect is negative, which means the flow through the key vertex tends to decrease when an arbitrary vertex is deleted.  This is to be expected because removing a vertex decreases the overall flow in the graph, which often causes the load on the key vertex to decrease.  However, as seen in the set of columns `Best \% Load Effect', in every single graph tested there existed at least one vertex whose load effect on the key vertex was positive.  In some cases, the best possible load effect on the key vertex is quite large. This demonstrates that optimal vertex deletion can generally increase the key vertex's load when the key vertex is highly central.

\subsection{Finding a needle in a haystack} \label{subsec:difficulties}

Having established that vertex deletion can force increased flow through a key vertex, we would like to characterize vertices having a positive load effect.  Although the brute force search for high load effect vertices can be done in polynomial time using standard maximum flow algorithms, this approach is problematic for two reasons: 1) It does not scale well to removing subsets of vertices rather than a single vertex, and 2) It does not offer any structural insights into the changes in flow routing that occur when a vertex is removed from a graph.  There are several obstacles to developing such a characterization, however.

The first obstacle is that vertices having a large positive load effect on the key vertex appear to be rare relative to the size of the graph, regardless of the graph type.  Table \ref{table:numpos} shows the average number of vertices having a positive load effect on the key vertex, by graph type, as well as the average number of vertices whose load effect is at least 75\% as large as that of the vertex with the largest load effect. In all but the Holme-Kim power law with clustering graphs, roughly 15-20\% of vertices have positive load in each graph type; in the Holme-Kim graphs only 7\% of vertices have positive load effect.  However, in all graph types, only 1-3 vertices, on average, in each graph have a load effect that is at least 75\% as large as the largest load effect. Thus, good vertices to delete are rare.    \noindent
\begin{table}
\begin{center}
{\scriptsize \begin{tabular}{|b{4cm}|c|c|c|c|c|c|}
\hline
 & \multicolumn{3}{c|}{Number of Vertices with } & \multicolumn{2}{c|}{Number of Vertices with } \\
 & \multicolumn{3}{c|}{Positive Load Effect} & \multicolumn{2}{c|}{Load Effect $\geq$ 75\% of Opt.} \\ \cline{2-6}
Graph Type & Min & Mean & Max & Mean & Std. Dev. \\
\hline
Erd\H{o}s-R\'{e}nyi random ($p=0.1$) &6 & 20.7 & 38 & 2.3 & 1.4\\
\hline
Watts-Strogatz small world ($p= 0.1, k=2$) & 4 & 23.4 & 59 & 2.8 & 2.2\\
\hline
Barab\'{a}si-Albert power law ($m_0=3, m=2$) & 3 & 15.1  & 47 & 1.5 & 0.8\\
\hline
Holme-Kim power law w. clustering ($m_0=2, m=2$) & 1 & 6.9  & 20 & 1.6 & 0.8\\
\hline
\end{tabular}\caption{Prevalence of positive load effect vertices. For each of four graph types we give statistics on the number of vertices per graph having positive load effect and the number of vertices per graph having load effect at least as large as 75\% of the largest load effect in the graph, over four samples of 276 100-vertex graphs of four types. \label{table:numpos}}}
\end{center}\end{table}

\begin{figure}[h!]
\begin{center}
\includegraphics[width=5in]{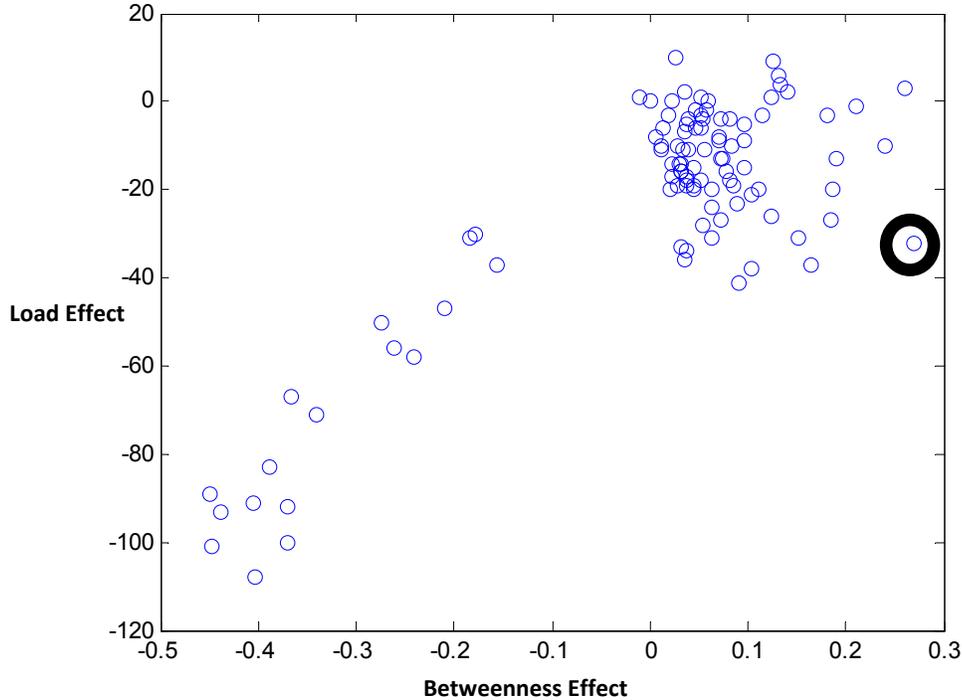}
\end{center}
\caption{Load effect versus betweenness effect in a 100-vertex random graph.  The circled point corresponds to the vertex having the largest betweenness effect on the key vertex; despite the positive correlation, we see that its load effect is far from maximum.}
\label{fig:HeuristicVar}
\end{figure}

As a result, simple heuristics that attempt to identify high load effect vertices based on correlations with other metrics are also unsuccessful.  For instance, we can define the \textit{betweenness effect of $i$ on $k$} to be the change in betweenness centrality of key vertex $k$ when vertex $i$ is removed.  Figure~\ref{fig:HeuristicVar}  shows a high correlation between betweenness effect and load effect, as we might expect.  However, if we were to target the highest betweenness effect vertex (circled), we would not select a vertex with a high load effect due to the variability about the trend line; in fact the load effect of this vertex is negative.  Similar problems arise when using other selection metrics, including structural equivalence\footnote{\textit{Structural equivalence} is the correlation between the binary vectors of neighbor sets of two vertices in a graph~\cite{WaF}.  If this correlation is high, then one expects the two vertices to have a similar role in the network.}.  Moreover, we have found that the choice of key vertex strongly influences the choice of the best vertex to target.  Heuristics that are based on generic properties of vertices in the graph ignore the relation between the key vertex and the targeted vertex.

Thus, any successful heuristic will likely need to leverage the specific structure of the network and the process by which it evolved and make explicit consideration of the key vertex to be able to hone in on the very small number of vertices worth targeting.

\subsection{Theoretical results} \label{subsec:proofs}
Although it is difficult to characterize properties of vertices having a high load effect on the key vertex, it is easier to rigorously demonstrate when a vertex is guaranteed \textit{not} to have a high load effect on the key vertex.  We present those results here.

The first observation is that the existence of cycles is critical to our method of removing vertices in order to maximize the load of a key vertex. The reason is that an increase in load implies that detours avoiding vertex $k$ have now been rerouted through $k$ upon deletion of another vertex.  Therefore, there must have been at least two edge-disjoint paths between some pair of vertices, one which passed through $k$ and one which did not.  This implies the existence of a cycle.  From this we can conclude that if the key vertex $k$ is a leaf, its load can never be increased through the removal of another vertex (and in fact, the load of a leaf is always equal to 0).  Therefore, our method of vertex deletion to increase load works only if the key vertex has degree at least 2.

For a similar reason, if the key vertex $k$ has degree exactly $2$, then removing vertices adjacent to $k$ will not increase $k$'s load, as the following theorem states:

\begin{theorem} \label{theorem:kDegree2}
Given a graph $G$ with key vertex $k$ having degree $2$, and vertex $i$ adjacent to $k$, $\mathcal{E}_k(G, \{i\}) \leq 0$.
\end{theorem}

By a similar logic, if the graph is a simple $n$-cycle ($n \geq 3$), then there is no vertex in the graph whose removal can increase the load on a key vertex $k$:
\begin{theorem} \label{theorem:nCycleLoad}
Let $G$ be a simple $n$-cycle.  Then for any choice of key vertex $k$ and vertex $i \neq k$ in $G$, $\mathcal{E}_k(G, \{i\}) \leq 0$.
\end{theorem}

Likewise, removing any vertex $i$ that does not lie on a same cycle as $k$ cannot increase the load on $k$ for there is no alternate path to reroute flow through $k$.  $i$ and $k$ do not lie on a same cycle when the maximum flow between them is equal to $1$, that is, there is only one edge-disjoint path between them.  We state this more formally as follows:
\begin{theorem} \label{theorem:onepath}
Let $k$ and $i$ be distinct vertices in graph $G$.  If there is only one edge-disjoint path between $k$ and $i$, then $\mathcal{E}_k(G, \{i\}) \leq 0$.
\end{theorem}

Theorem \ref{theorem:onepath} is actually a special case of a general theorem related to the size of a cut between the key vertex $k$ and a candidate for removal, $i$, which we present here:

\begin{theorem} \label{theorem:nEDpaths}
Let $k$ and $i$ be distinct vertices in graph $G$.  Consider an edge cut $C$ that partitions $G$ into two components such that $i$ and $k$ are in separate components.  Let $G_k$ be the subgraph of $G$ over the set of vertices in the component containing $k$, and let $G_i$ be the subgraph of $G$ over the set of vertices in the component containing $i$.  Let $i_1,...,i_p$ be the vertices on the $i$ side of the cut that are adjacent to $G_k$.  Let $k_1,...,k_s$ be the vertices on the $k$ side of the cut adjacent to $G_i$.  Suppose any boundary vertex $i_1 \in G_i$ has at least $\lfloor |C|/2 \rfloor$ edge-disjoint paths to every other boundary vertex of $G_i$ by using only vertices in $G_i\setminus \{i\}$, and any boundary vertex $k_1 \in G_k$ has at least $\lfloor |C|/2 \rfloor$ edge-disjoint paths to every other boundary vertex of $G_k$ by using only vertices in $G_k\setminus \{k\}$.  Then $\mathcal{E}_k(G, \{i\}) \leq 0$.
\end{theorem}

\begin{figure}[h]
	\centering
	\qquad
	\subfloat[][Graph $G$]{
	\includegraphics[width=3in]{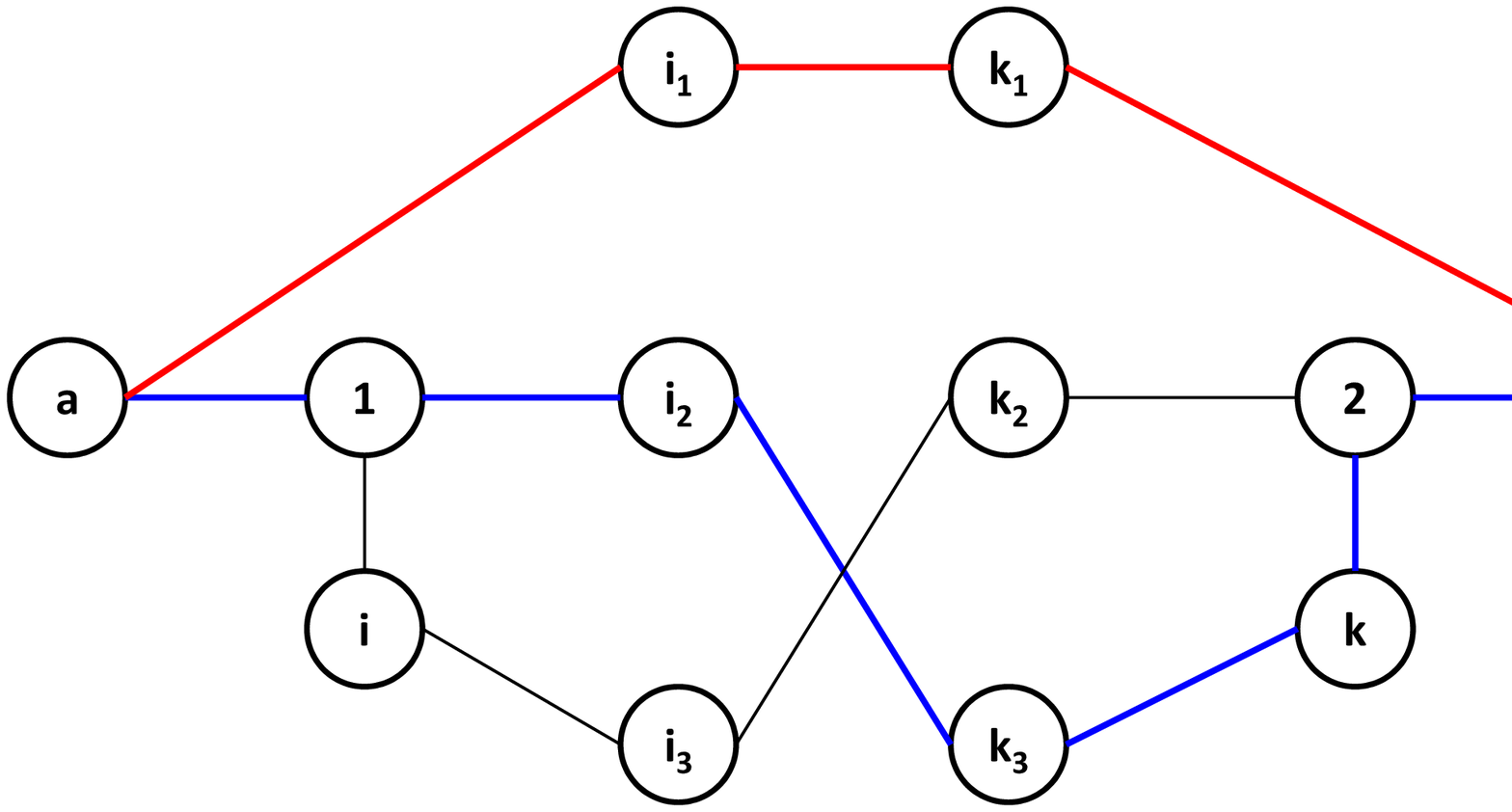}
	\label{fig:ProofFig1_G}}
	\subfloat[][Graph $G \setminus \{k\}$]{
	\includegraphics[width=3in]{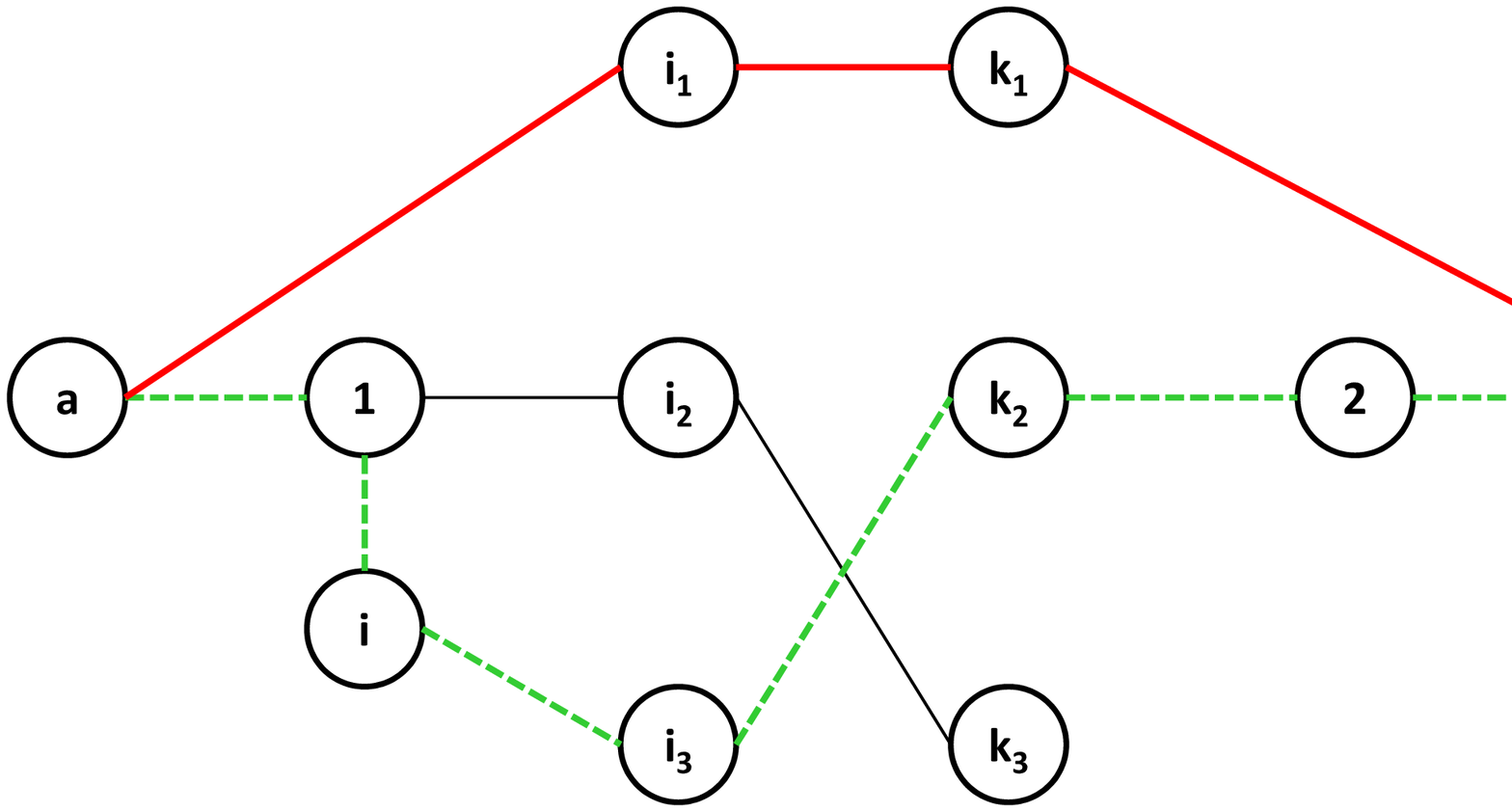}
	\label{fig:ProofFig1_G_k}}
	\qquad
	\subfloat[][Graph $G \setminus \{i\}$]{
	\includegraphics[width=3in]{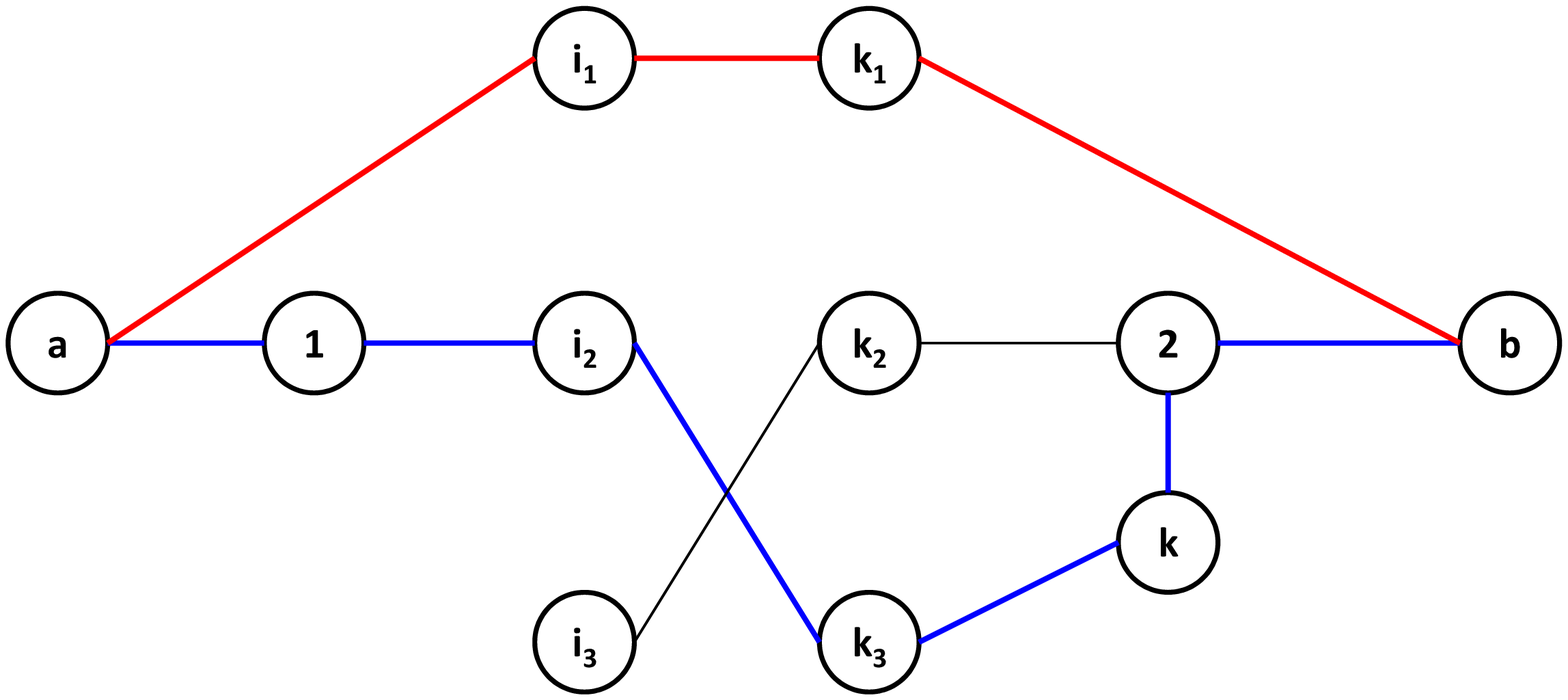}
	\label{fig:ProofFig1_G_i}}
	\subfloat[][Graph $G \setminus \{i,k\}$]{
	\includegraphics[width=3in]{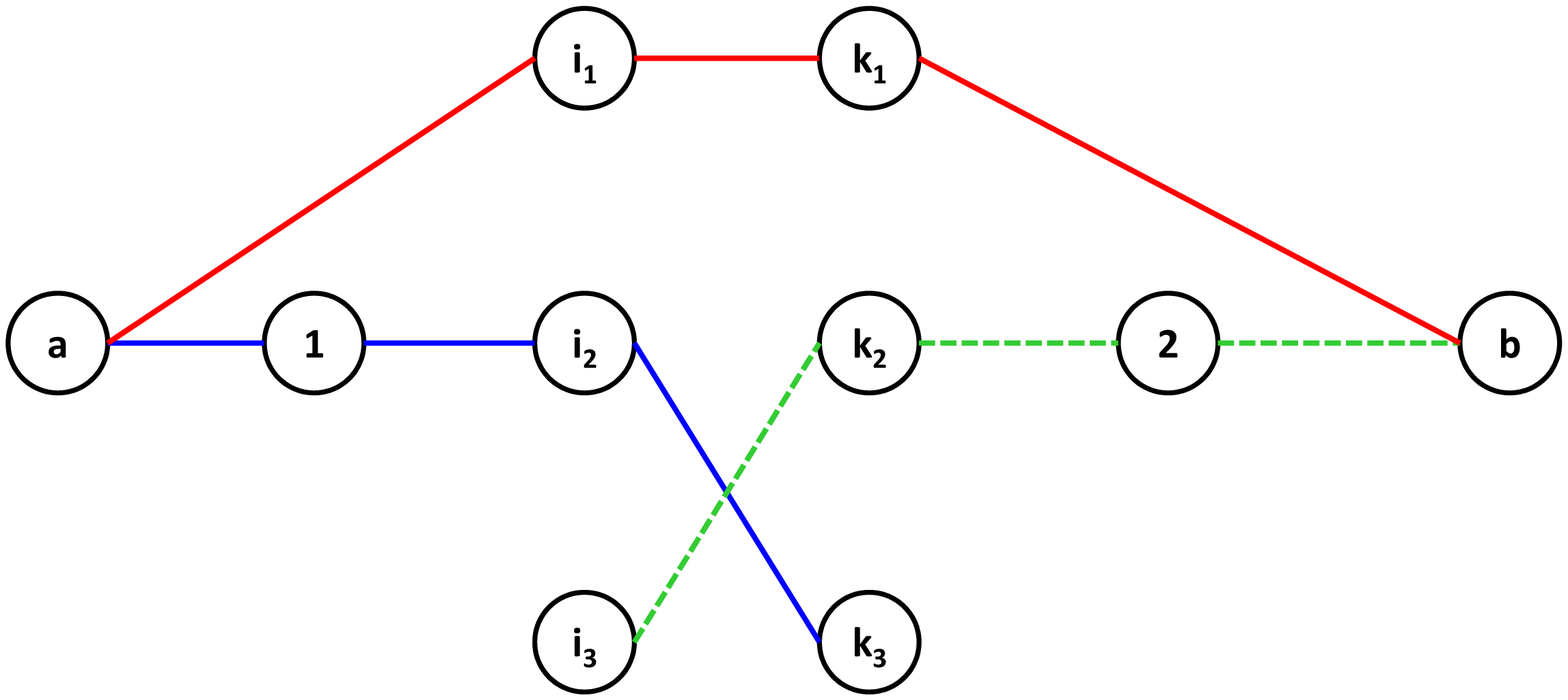}
	\label{fig:ProofFig1_G_ik}}
	\caption{This example illustrates the contrapositive of Theorem \ref{theorem:nEDpaths}.  Removing vertex $i$ can increase the amount of $a-b$ flow passing through $k$, because there are not at least $\lfloor |C|\rfloor/2$ edge-disjoint paths between boundary vertices on each side of the cut that avoid $i$ and $k$.  Figure \subref{fig:ProofFig1_G} shows the original graph $G$ having a cut of size 3 between vertices $i$ and $k$.  The number of edge-disjoint paths between $a$ and $b$ in graph $G$ is 2.  Figure \subref{fig:ProofFig1_G_k} shows the graph with $k$ removed.  There are still two edge-disjoint paths between $a$ and $b$ in graph $G \setminus \{k\}$, so the load on $k$ in graph $G$ with respect to $a - b$ flow is zero.  Figure \subref{fig:ProofFig1_G_i} shows that when $i$ is removed, there are still two edge-disjoint paths between $a$ and $b$, and Figure \subref{fig:ProofFig1_G_ik} shows that when both $i$ and $k$ are removed, the number of edge-disjoint $a-b$ paths drops to one. Therefore, the load on $k$ with respect to $a - b$ flow in $G \setminus \{i\}$ is one, and the load effect on $k$ of removing vertex $i$ (with respect to flow between vertices $a$ and $b$) is positive.}
	\label{fig:ProofFig1}
	\end{figure}

\begin{figure}[h]
	\centering
	\qquad
	\subfloat[][Graph $G$]{
	\includegraphics[width=3in]{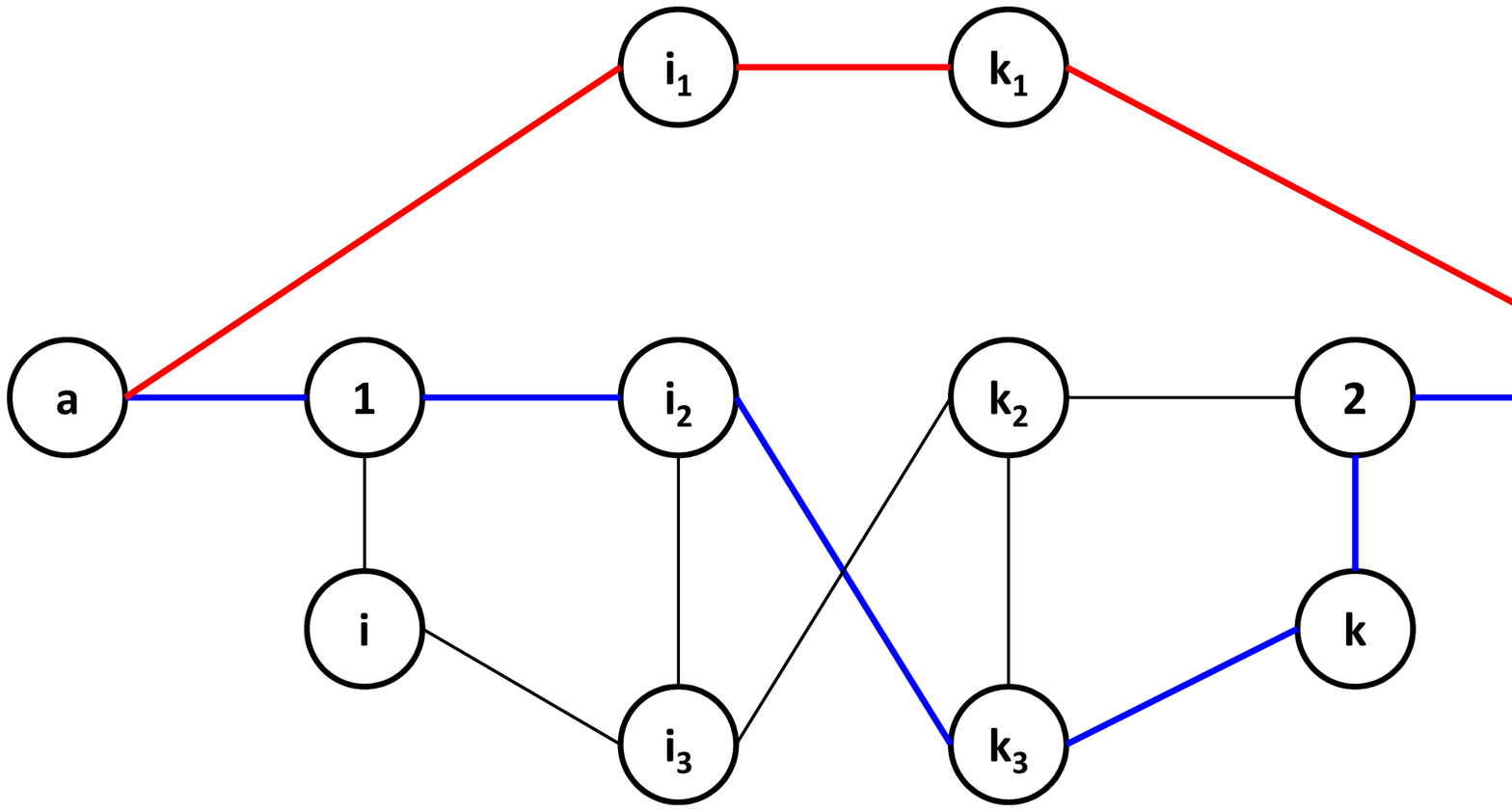}
	\label{fig:ProofFig2_G}}
	\subfloat[][Graph $G \setminus \{k\}$]{
	\includegraphics[width=3in]{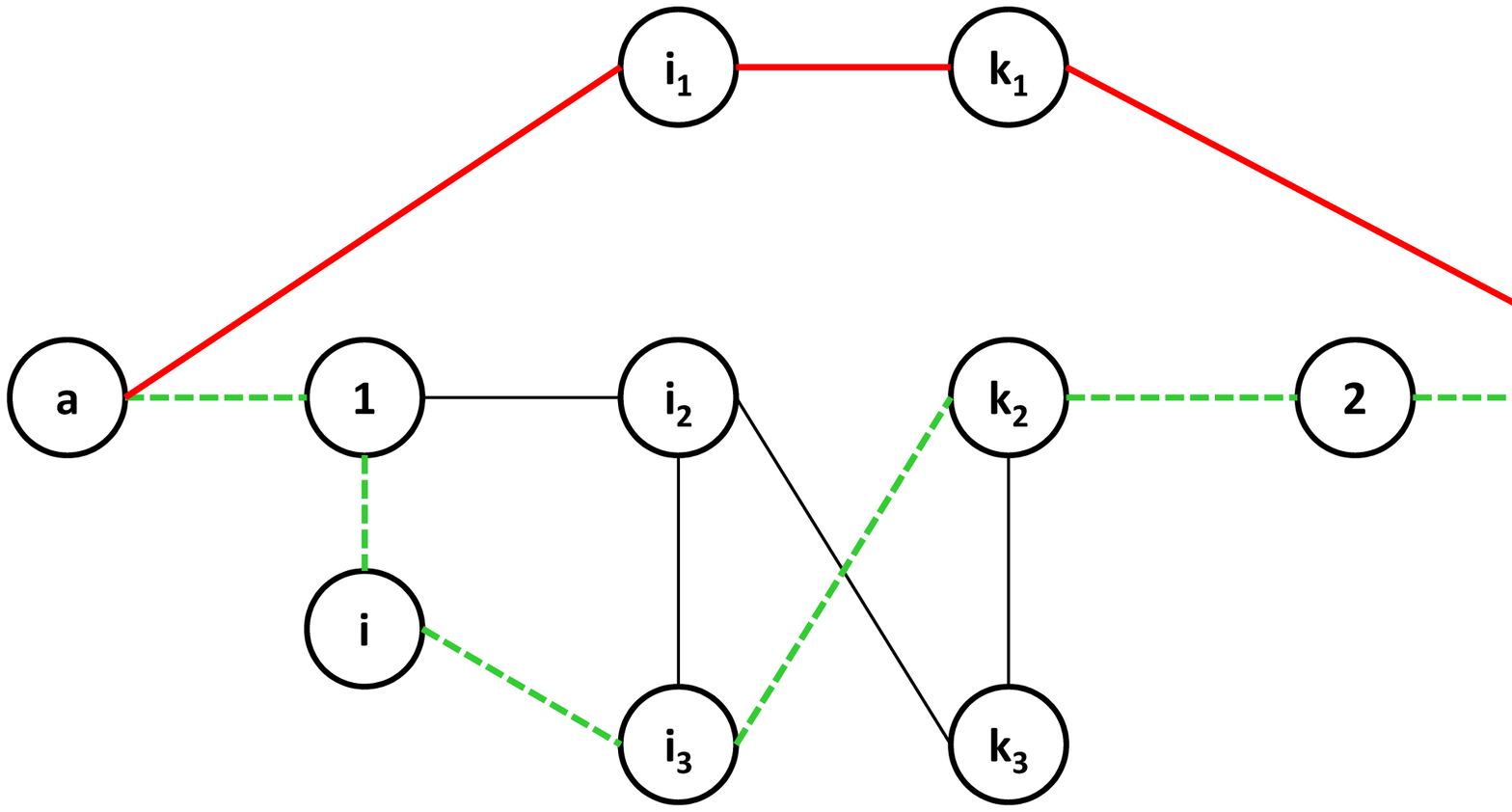}
	\label{fig:ProofFig2_G_k}}
	\qquad
	\subfloat[][Graph $G \setminus \{i\}$]{
	\includegraphics[width=3in]{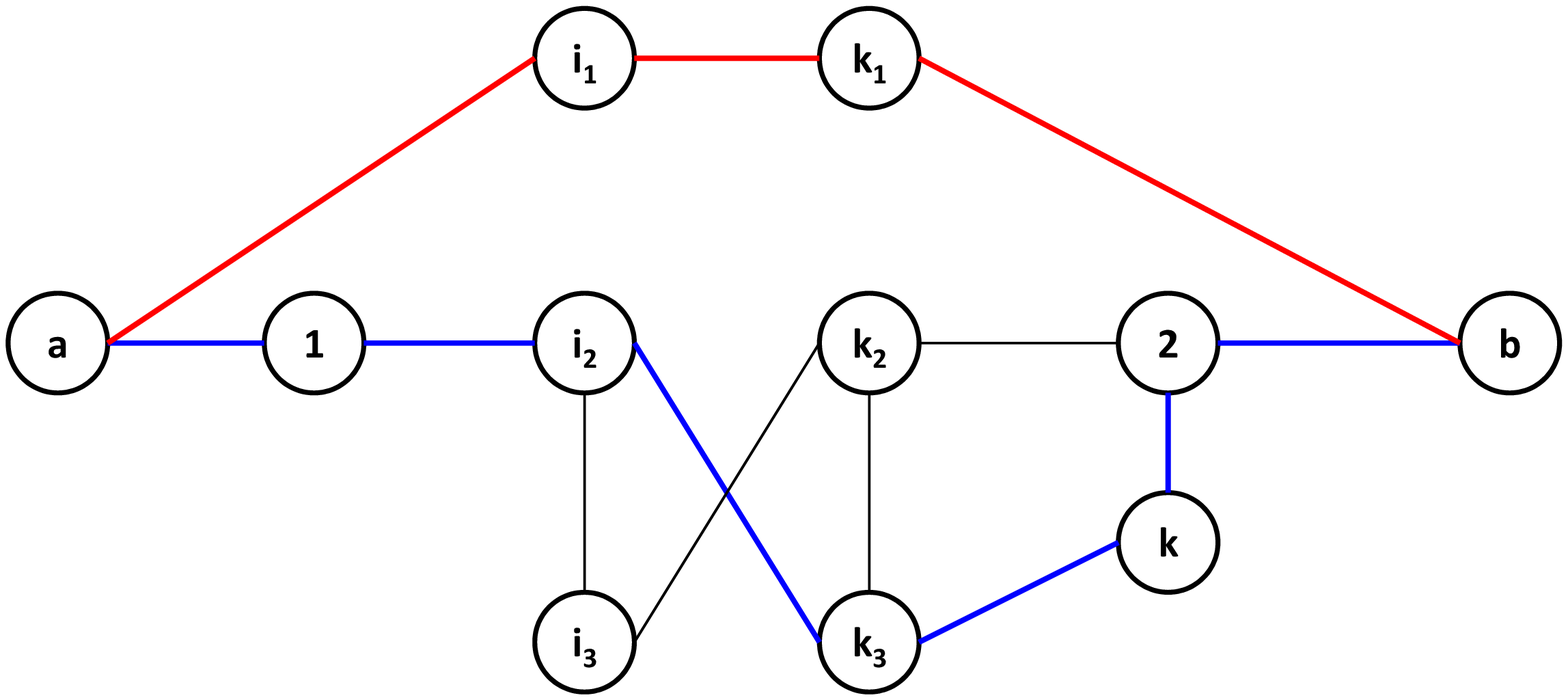}
	\label{fig:ProofFig2_G_i}}
	\subfloat[][Graph $G \setminus \{i,k\}$]{
	\includegraphics[width=3in]{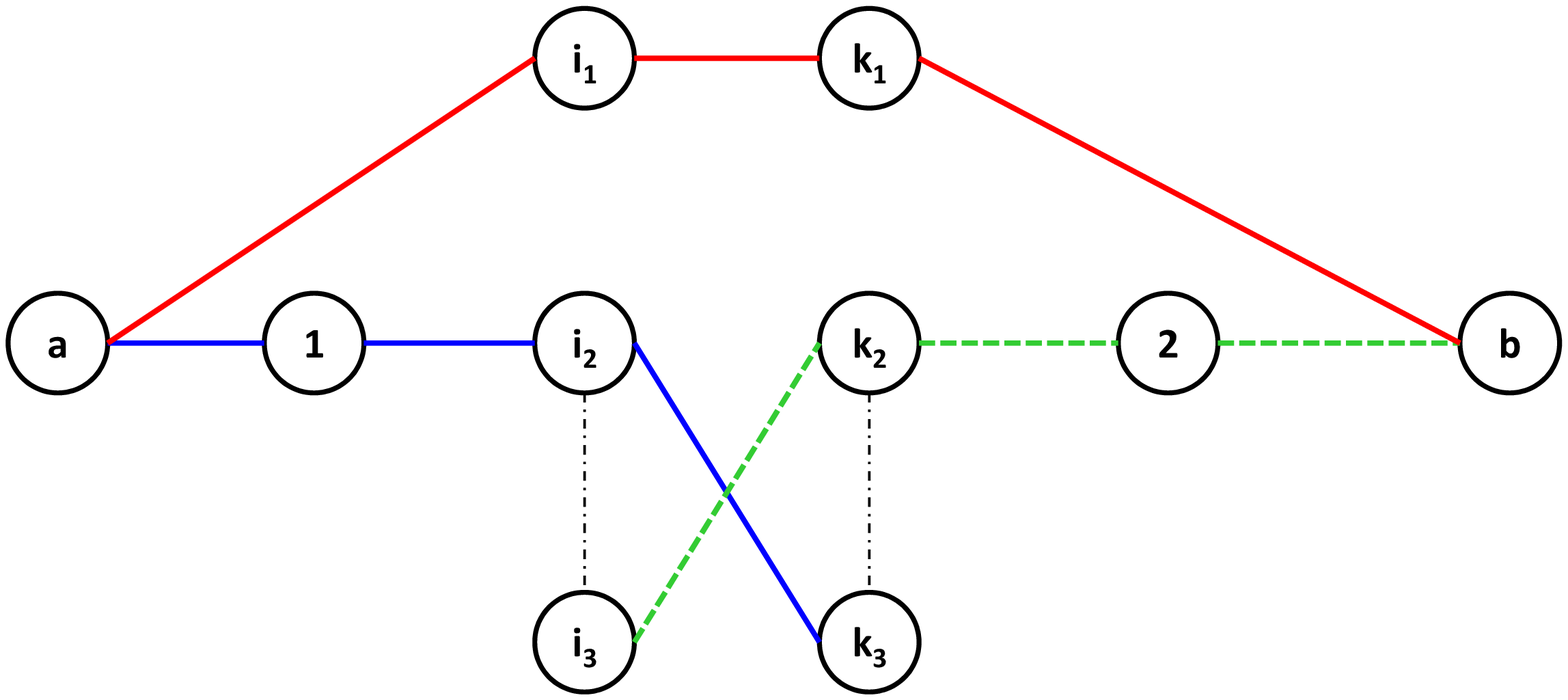}
	\label{fig:ProofFig2_G_ik}}
	\caption{This figure is identical to Figure \ref{fig:ProofFig1} but with additional edges added between boundary vertices in order to satisfy the conditions of Theorem \ref{theorem:nEDpaths}.  We see that the load effect on $k$ of removing vertex $i$ (with respect to flow between vertices $a$ and $b$) is now zero.  Figure \subref{fig:ProofFig2_G} shows the original graph $G$ having a cut of size 3 between vertices $i$ and $k$.  The number of edge-disjoint paths between $a$ and $b$ in graph $G$ is 2.  Figure \subref{fig:ProofFig2_G_k} shows the graph with $k$ removed.  There are still two edge-disjoint paths between $a$ and $b$ in graph $G \setminus \{k\}$, so the load on $k$ with respect to $a - b$ flow in graph $G$ is zero.  Figures \subref{fig:ProofFig2_G_i} and \subref{fig:ProofFig2_G_ik} show that even when $i$ is removed, the load on $k$ with respect to $a - b$ flow in graph $G$ is still zero.  This is because the existence of edge-disjoint paths between boundary vertices guarantees that the two half-paths from Figure \ref{fig:ProofFig1_G_ik} can now be stitched together to maintain a unit of flow even when $k$ is removed.  }
	\label{fig:ProofFig2}
	\end{figure}

The general idea of this theorem is that a vertex $i$ cannot increase the load on key vertex $k$ if there exist too many detours across the cut between it and $k$.  This is depicted in Figures \ref{fig:ProofFig1} and \ref{fig:ProofFig2}.  In Figure \ref{fig:ProofFig1_G} we have a graph $G$ that has a cut separating $i$ and $k$ having boundary vertices $i_1$, $i_2$, and $i_3$ on the $i$ side of the cut and $k_1$, $k_2$ and $k_3$ on the $k$ side of the cut.  The cut has capacity three, and so we require at least one edge-disjoint path that avoids $i$ and $k$ between each pair of boundary vertices on each side of the cut.  We see that this requirement is not met, as vertices $i_2$ and $i_3$ have no path between them on the $i$ side of the cut that avoids $i$, and vertices $k_2$ and $k_3$ have no path between on the $k$ side of the cut them that avoids $k$.  Moreover, we see that there are two edge-disjoint paths between $a$ and $b$ in graphs $G$, $G \setminus \{k\}$, and $G \setminus \{i\}$ but only one edge-disjoint path in $G \setminus \{i,k\}$.  Therefore, the load effect on $k$ of removing vertex $i$ with respect only to flow between $a$ and $b$ is 1, which is positive.  We lose a unit of flow because there is no way to connect the blue half-path with the green dotted half-path of Figure \ref{fig:ProofFig1_G_ik}. Figure \ref{fig:ProofFig2_G} gives the same graph but with extra edges added to satisfy the theorem conditions.  Now we see that the load effect of $i$ on $k$ is 0 because of the presence of detours that allow paths to avoid $k$ even after $i$ has been removed from the graph.  The blue and green half-paths of Figure \ref{fig:ProofFig2_G_ik} can be connected using edge-disjoint paths between boundary vertices so that the path from $a$ to $b$ can remain complete.

The proofs of these theorems can be found in the Appendix.  We can use these theorems within a heuristic for reducing the number of vertices explored in a brute force computation of load effects.  This is described in the next section.

 \subsection{Heuristics} \label{subsec:heuristics}

\noindent \textbf{Vertex elimination heuristic}

\noindent In order to use the theorems of the previous section to rule out vertices, we must be able to identify vertices having the properties in the theorem statements.

We can ignore Theorem \ref{theorem:nCycleLoad} because rarely will our graph of interest be a cycle.  Theorem \ref{theorem:kDegree2} is straightforward to implement:  Given the adjacency matrix for the graph, we can sum the row corresponding to $k$ in $\mathcal{O}(|V|)$ time to determine if $k$ has degree 2.  If it does, then we can ignore its neighbors in our brute force computation.  Although this will save us at most 2 load computations, each load computation requires two all-pairs maximum flow computations, and the algorithm for solving the all-pairs maximum flow problem in practice takes $\mathcal{O}(|V|^3\sqrt{|E|})$ time using a Gomory-Hu tree \cite{CM1999, GH1961}.

Theorem \ref{theorem:onepath} is likewise straightforward to implement.  Prior to starting the brute force search, it is necessary to compute the load on the key vertex in the original graph.  This requires computation of the all-pairs maximum flow in the original graph.  We can therefore use the resulting Gomory-Hu tree from the original graph to identify, in $\mathcal{O}(|V|)$ time, the vertices having only one edge-disjoint path to the key vertex.

Theorem \ref{theorem:nEDpaths} cannot be fully implemented to identify all vertices with non-positive load because enumerating every possible cut in the graph takes far longer than computing every load effect.  However, used in moderation, the theorem can help prune out some vertices whose load effect cannot be positive.  We again assume we have a Gomory-Hu tree for the original graph $G$ and the associated set of flow paths (attainable from the Gusfield implementation of the Gomory-Hu algorithm \cite{G1990}).  The $n-1$ edges of this tree identify $n-1$ cuts.  For each cut $C$, we can identify the boundary vertices of the cut (and we ignore any cut for which $k$ is already on the boundary), and define the two sides $G_i$ and $G_k$ depending on the location of vertex $k$.  We select a vertex on the boundary of $G_k$ and compute the maximum flow between it and all other boundary vertices using only edges in $G_k \setminus \{k\}$.  Likewise, we select a vertex on the boundary of $G_i$ and compute the maximum flow between it and all other boundary vertices using only edges in $G_i$.  If any of these flow values are smaller than $\lfloor |C|/2 \rfloor$, then our theorem condition is not met, we discard our current cut $C$ and we move on to the next cut in the Gomory-Hu tree.

However, if all of these flow values are at least $\lfloor |C|/2 \rfloor$, then we initialize a set $R$ consisting of all non-boundary vertices in $G_i$.  As the heuristic proceeds, we will remove from this set any vertex whose removal results in fewer than $\lfloor |C|/2 \rfloor$ paths between any boundary pair.  At the end of the heuristic, any vertex remaining in the set $R$ will satisfy the conditions of the theorem statement and will be known to have a non-positive load effect on $k$.  We identify these vertices by examining the flow paths between boundary vertices:
  \begin{itemize}
  \item Any vertex that does not lie on any flow path between any pair of boundary vertices in $G_i$ does not affect the passage of flow between boundary vertices if removed from the graph.  Therefore any such vertex satisfies the theorem condition, remains in the set $R$, and is known to have a non-positive load effect.
  \item For each vertex $i$ that appears on a flow path between a pair of boundary vertices, we compute the flow between boundary vertices over the graph $G_i \setminus \{i\}$.
      \begin{itemize}
      \item If the flow between every pair of boundary vertices remains at least $\lfloor |C|/2 \rfloor$, then $i$ satisfies the theorem condition, remains in the set $R$, and is known to have a non-positive load effect.
      \item  If the flow between any pair of boundary vertices drops below $\lfloor |C|/2 \rfloor$, then it is possible that $i$ could have a positive load effect on $k$.  It is removed from the set $R$ and, as long as it is not added to the set $R$ upon examination of a future cut, it will eventually be evaluated during the brute force computation of load effects.
          \end{itemize}
          \end{itemize}
  The vertices remaining in the set $R$ at the end of this process are permanently flagged as having non-positive load effects on $k$ and need never be considered by any future cuts.

Clearly, this heuristic has no guarantee of identifying every vertex having a non-positive load effect in the graph.  Moreover, it is possible that the heuristic could evaluate every vertex in the graph, which would be almost as difficult as calculating the load effect directly.  However, as long as only small cuts in the Gomory-Hu tree are explored, the heuristic decreases the overall computation time required to perform a brute force evaluation of all vertices' load effects, as we now discuss.

We examined cuts of sizes one through five, as cuts of greater capacity rarely have any vertices satisfying the theorem statement.  In the four graph types discussed earlier, the heuristic identifies on average less than one vertex (Table~\ref{results}). We attribute the failure of this heuristic to the overall connectedness of the graphs used in our testbed.

This heuristic is most successful on graphs that have clusters of highly connected components that are only loosely connected to one another.  An example of such a network is the Global Salafi Jihad terrorist network that has four clusters that are loosely connected to each other: the Central Staff of al Qaeda, the Maghreb Arab Cluster, the Core Arab Cluster and the Jemaah Islamiyah \cite{Sag:2004}.  The four clusters are loosely connected to each other via connections between leaders from each group.  The four clusters themselves are more densely connected.   While we see this behavior in real world terrorist networks, it is not exhibited in the randomly generated graph types mentioned earlier.  However, we can construct graphs having a similar structure; we call this graph type \textit{centralized power law} graphs.  We start by creating a ``leadership group'' using the Barab\'{a}si-Albert preferential attachment generation method.  Some vertices in this leadership group are then also assigned as leaders to satellite groups; each satellite group must have at least one leader.  Then new vertices are generated and are connected to vertices within only one satellite group, using the same preferential attachment scheme.  The number of edges attached to a new vertex is randomly chosen from a uniform distribution from 1 to a user specified number.  This means that new vertices are added with varying degrees of connectedness.  Thus, the potential for leaves still remains, while an asymptotically power law degree distribution is maintained.

\begin{figure}[h]
\begin{center}
\includegraphics[width = 5in]{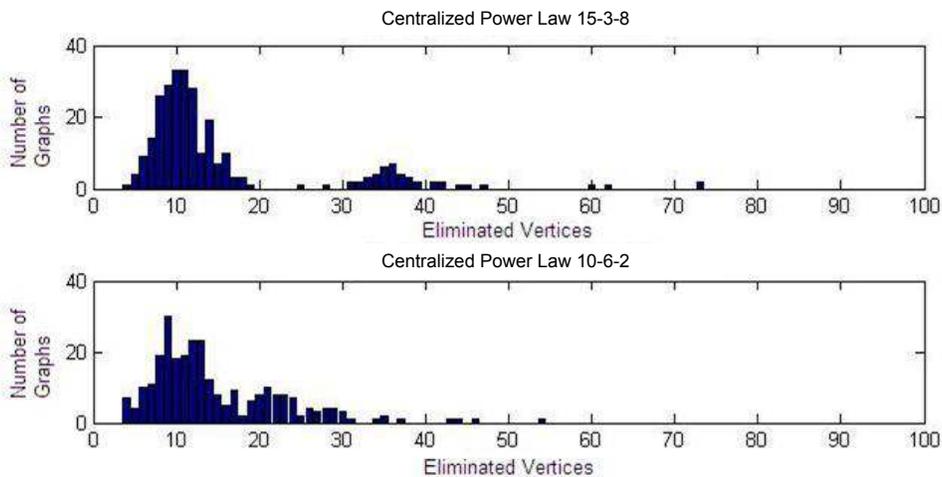}
\caption{Distribution of the number of vertices identified by the vertex elimination heuristic as having non-positive load effect on the key vertex.  The two graph types presented are centralized power law graphs with parameters $a-b-c$, where $a$ is the number of members in the leader group, $b$ is the number of satellite groups, $c$ is the number of leaders not assigned to any satellite group.}\label{Hist}
\end{center}
\end{figure}

Table~\ref{results} presents the average number of vertices having non-positive load effects that were identified by the heuristic in each graph of that type; the number of vertices the heuristic would need to identify in order to save time on a brute force computation; the average number of vertices in each graph having a negative load effect (as determined by our brute force results mentioned earlier); and the computation required to run the heuristic plus the subsequent brute force load effect computation on the remaining vertices, expressed as a percentage of the original brute force run time.  On the centralized power law graphs, the heuristic performs well.  Figure~\ref{Hist} presents a histogram of the distribution of the number of vertices identified by the heuristic over the 276 graph instances for each of two types of centralized power law graphs.

 The number of vertices that the heuristic needs to remove from consideration in order to reduce the total runtime of the brute force load effect computation is between 2 and 6.  On the types of graphs where the heuristic identifies many vertices, it saves up to 12\% of the runtime.  On those graph types where less than one vertex is identified on average, the heuristic adds up to 5\% to the runtime.  Therefore, it is a useful preprocessor for brute force computation of load effects on graph types having sparsely interconnected clusters, but less useful on more dense graphs.

\begin{table}
\begin{center}
{\scriptsize \begin{tabular}{|b{4cm}|b{1.5cm}|b{2.5cm}|b{1.5cm}|b{2cm}|}
\hline
Graph Type  & Avg. Number Identified  & Avg. Number with Non-positive  Load Effect      &    Number Required   & \% Brute Force Total Time\\
\hline
Erd\H{o}s-R\'{e}nyi random graph  &&&& \\
($p = 0.1$) & 0.04 & 79.3& 2 & 101.54 \\
\hline
Watts-Strogatz small world &&&& \\
($p = 0.1, k=2$) & 0.63 & 76.6 & 5 & 104.16 \\
\hline
Barab\'{a}si-Albert &&&& \\
power law &&&& \\
($m_0=3, m=2$) & 0.08 & 89.6 & 6 & 105.02 \\
\hline
Holme-Kim power law &&&& \\
  w. clustering&&&& \\
($m_0=2, m=2$) & 0.02 & 87.5 & 2 & 101.58 \\
\hline
Centralized power law &  & &  &  \\
15 leaders, 3 groups &14.70 & 98.2&4 & 88.8 \\
\hline
Centralized power law & & & & \\
10 leaders, 6 groups & 15.40 & 93.9 & 4 & 88.01 \\
\hline
\end{tabular}
\caption{Runtime and effectiveness of the vertex elimination heuristic on six graph types:  The average number of vertices identified by the vertex elimination heuristic as having non-positive load effects; the average number of vertices in the graph having non-positive load effects; the number of vertices that would need to be identified by the heuristic to improve the running time of brute force identification; the total time required to run the heuristic and brute force computation, as a percentage of the original brute force computation time. \label{results}}}
\end{center}
\end{table}

\noindent \textbf{Divide and conquer heuristic}

We also developed a divide and conquer heuristic that typically finds a good solution.  The intuition behind our heuristic is that if a subset of vertices has a high load effect, then that subset may contain an individual vertex with a high load effect.   The heuristic starts by calculating the load effects of  $s$ different subsets of vertices of size $t$ that partition the set $V$.  Subsets having a large positive load effect become candidates for further investigation in which the load effect of removing each member of the subset individually is computed. Our heuristic partitions the graph into $s$ equally-sized subsets of vertices and calculates the load effect of each subset. Once the $t$ subsets with the highest load effects have been identified, the load effect of every vertex within these $t$ subsets is computed and the best one is selected. In a graph of $|V|$ vertices, partitioned into $s$ subsets of $k$ vertices each ($sk=|V|$), the divide and conquer strategy will compute only $s+tk$ load effects rather than the $|V|=sk$ load effects calculated in a brute force attempt.

Table \ref{table:divConq} shows the results of using the divide and conquer heuristic on 100-vertex graphs whose vertex sets were partitioned into subsets of size five, and the top $t=1$, $t=2$, and $t=3$ subsets were explored fully. The first, fourth and seventh columns of data show the load effect of the best vertex identified by the heuristic, presented as a percentage relative to the known best load effect in the graph, averaged over 276 graphs.  We also show the average ranking of the load effects found by the heuristic (data columns two, five and eight), and the fraction of trials where the vertex identified by the heuristic had a \textit{negative} load effect (data columns three, six and nine).   We see that the divide and conquer approach is effective at identifying vertices that have large positive load effects on the key vertex, and on average identifies the top 1 to 3 vertices in the graph when as few as two top subsets ($t=2$) are investigated.

Both methods fail to provide us with an understanding of the structural characteristics of the best vertices to target.  This remains  a key question for the Single-LOMAX problem.

\noindent \begin{table} \begin{center}{\scriptsize
\begin{tabular}{|b{3cm}|c|c|c|c|c|c|c|c|c|c|c|c|}
\hline
 & \multicolumn{3}{c|}{Best Subset ($t=1$)} & \multicolumn{3}{c|}{Best 2 Subsets ($t=2$)} & \multicolumn{3}{c|}{Best 3 Subsets ($t=3$)} \\
 \cline{2-10}
\multirow{2}{*}{Graph Type} &\% of Best& Avg. &  \% & \% of Best& Avg. &  \% &  \% of Best& Avg. &  \% \\
 & Effect & Rank & Neg. & Effect & Rank & Neg. &  Effect & Rank & Neg. \\
\hline
Erd\H{o}s-R\'{e}nyi random  &&&&&&&&& \\
($p = 0.1$) & 62.4 & 4.9 & 1.8 &75.1 & 3.2 & 0.0 &  81.0 & 2.4 & 0.0\\
\hline
Watts-Strogatz small world &&&&&&&&&  \\
($p = 0.1, k=2$)&  66.0 & 6.1  & 7.6  & 78.4 & 3.1 & 1.8 & 83.6 & 2.2 & 0.0\\
\hline
Barab\'{a}si-Albert &&&&&&&&&  \\
power law &&&&&&&&&  \\
($m_0=3, m=2$) & 62.6 & 3.3 & 2.9  & 74.6& 2.2 & 1.1  & 78.2 & 1.9 & 0.3\\
\hline
Holme-Kim power law &&&&&&&&&  \\
  w. clustering&&&&&&&&&  \\
($m_0=2, m=2$) & 36.5 & 4.5 & 18.8  & 52.4 & 3.1 & 10.9 & 67.0 & 2.2 & 4.0\\
\hline
\end{tabular}
\caption{Performance of the divide and conquer method using subsets of size five and fully exploring the best $t=1$, $t=2$ and $t=3$ subsets.  The load effect of the best vertex identified by the heuristic is presented as a percentage relative to the known best load effect in the graph, averaged over 276 graphs.  We also show the average ranking of the load effects found by the heuristic, and the fraction of trials where the vertex identified by the heuristic had a \textit{negative} load effect.\label{table:divConq}}}
\end{center}
\end{table}

\section{Multiple vertex deletion (LOMAX)} \label{sec:multiple}
For graphs with a few hundred vertices, it is not unreasonable to perform a brute force computation to solve Single-LOMAX.  However, removing only a single vertex might not maximize the load on the key vertex; we might instead prefer to remove a subset of vertices from the graph in order to reroute flow through the key vertex.  We call this problem LOMAX, and in this case, it is computationally intractable to compute the load effect of every possible subset in the graph, and we must resort to other methods, as we describe in this section.

\subsection{Genetic algorithm}

Because load effect does not change smoothly according to a known function over subsets, LOMAX is not amenable to standard optimization techniques.  Instead, we designed a genetic algorithm to rapidly compute good quality solutions.  The genetic algorithm iteratively hybridizes good solutions to construct even better solutions. In this algorithm, we start with an initial \textit{solution pool}, which is a collection of candidate subsets of vertices to remove from the graph.  We will call the initial solution pool $P_0$, and we let $P_i$ denote the solution pool at the beginning of iteration $i$. During each iteration, the algorithm takes the following general steps:

\begin{enumerate}

\item \textit{Selection}. Partition $P_i$ into two equally sized sets, $P_i^{good}$ and $P_i^{bad}$, where the load effect of every solution in $P_i^{good}$ is greater than or equal to that of each solution in $P_i^{bad}$.

\item \textit{Recombination}. Recombine the solutions in $P_i^{good}$ to create (without loss of generality) $\frac{|P_i|}{2}$ new solutions.  For example, a solution comprised of vertices $\{i_1, i_2, \ldots, i_k\}$ can be recombined with the solution $\{j_1, j_2, \ldots, j_k\}$ to create a new solution $\{i_1, j_2, \dots, i_{k-1}, j_k\}.$ We denote the set of these new solutions with $P_{i+1}^R$.

\item \textit{Generation}. Randomly generate (without loss of generality) $\frac{|P_i|}{2}$ solutions from scratch. We denote the set of these new solutions with $P_{i+1}^N$.

\item \textit{Evaluation}. $P_{i+1} = P_{i+1}^R \cup P_{i+1}^N$ is the solution pool for the $(i+1)^{st}$ iteration. Compute the load effect of each solution in $P_{i+1}$. If any of these solutions has a better load effect than the best solution seen so far, save the new solution with the largest load effect as the new best solution.

\item \textit{Iterate}.

\end{enumerate}

At the end of each iteration, the algorithm checks if either of the two termination conditions are satisfied. Our algorithm can terminate if either the number of iterations performed exceeds a user-specified number, or the number of iterations where a better solution is not found exceeds a user-specified number.

This is the general framework for our genetic algorithm.  In the next section, we describe the specifications of our own implementation of the algorithm and our computational results.

\subsection{Computational results}

We implemented the algorithm using a solution pool of size twenty, with each solution in the pool corresponding to a subset of up to six vertices.  Subsets of size smaller than six were represented using dummy vertices that could also be swapped during Recombination and chosen randomly during Generation.  The greater the likelihood assigned to dummy vertices, the more likely a solution would have fewer than six vertices; we assigned a dummy vertex the same selection probability as all other vertices. At each iteration, ten new solutions were created during Recombination by hybridizing the current best ten solutions, and ten solutions were created randomly during Generation to replace the current worst ten solutions.  During Recombination, there are many possible ways to divide each solution in half, and four ways to hybridize any pair of solutions.  In our implementation, we divided a solution in half based on the order in which the vertices are listed in the data structure; this is kept unsorted to avoid biasing the recombinations.  We then examined the four possible hybridizations and picked the first that had not previously been chosen and had no vertices duplicated.  If none of the four hybridizations satisfied these two conditions, then we skipped the hybridization of these two solutions and instead generated a solution at random.

\begin{figure}
	\centering
	\includegraphics[width=5in]{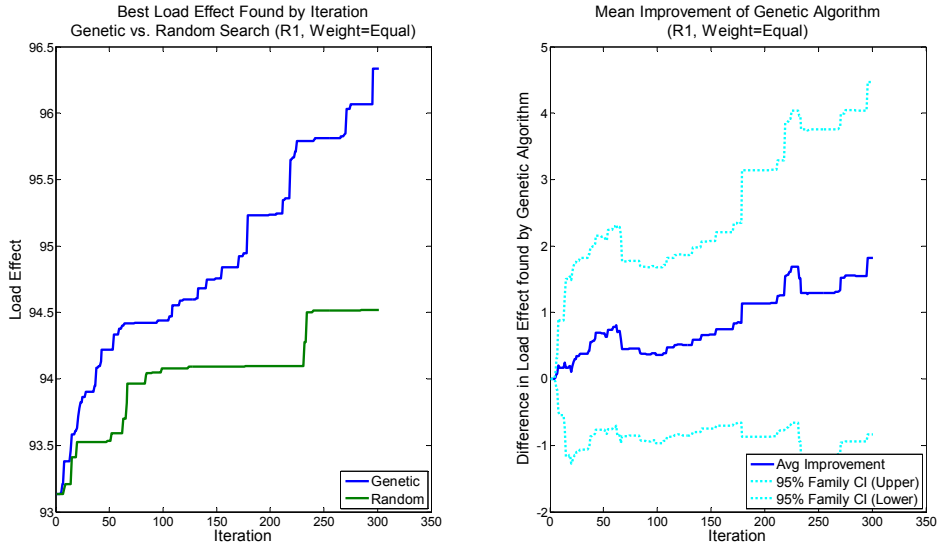}
	\caption{Performance of the genetic algorithm as compared to random search on 276 instances of Erd\H{o}s-R\'{e}nyi random graphs. \label{fig:R1GenAlg}}
	\end{figure}

\begin{figure}[h]
	\centering
	\includegraphics[width=5in]{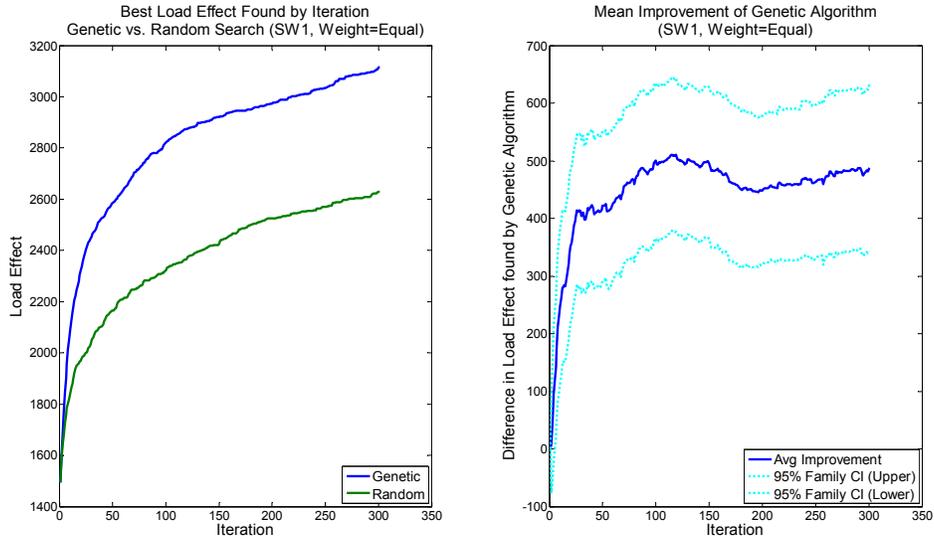}
	\caption{Performance of the genetic algorithm as compared to random search on 276 instances of Watts-Strogatz small world graphs. \label{fig:SW1GenAlg}}
	\end{figure}

\begin{figure}[h]
	\centering
	\includegraphics[width=5in]{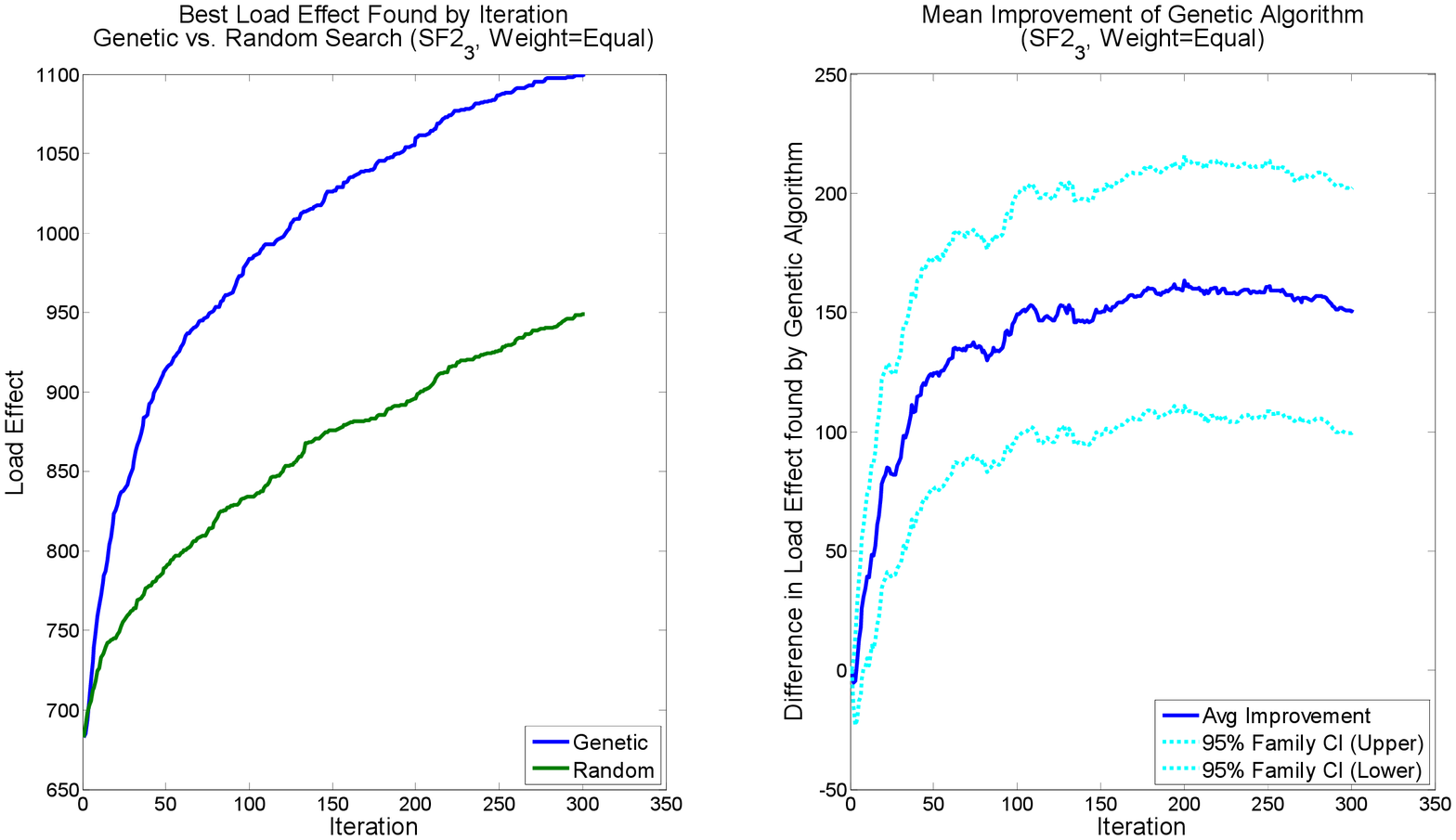}
	\caption{Performance of the genetic algorithm as compared to random search on 276 instances of Barab\'{a}si-Albert power law graphs. \label{fig:SF2_3GenAlg}}
	\end{figure}

\begin{figure}[h]
	\centering
	\includegraphics[width=5in]{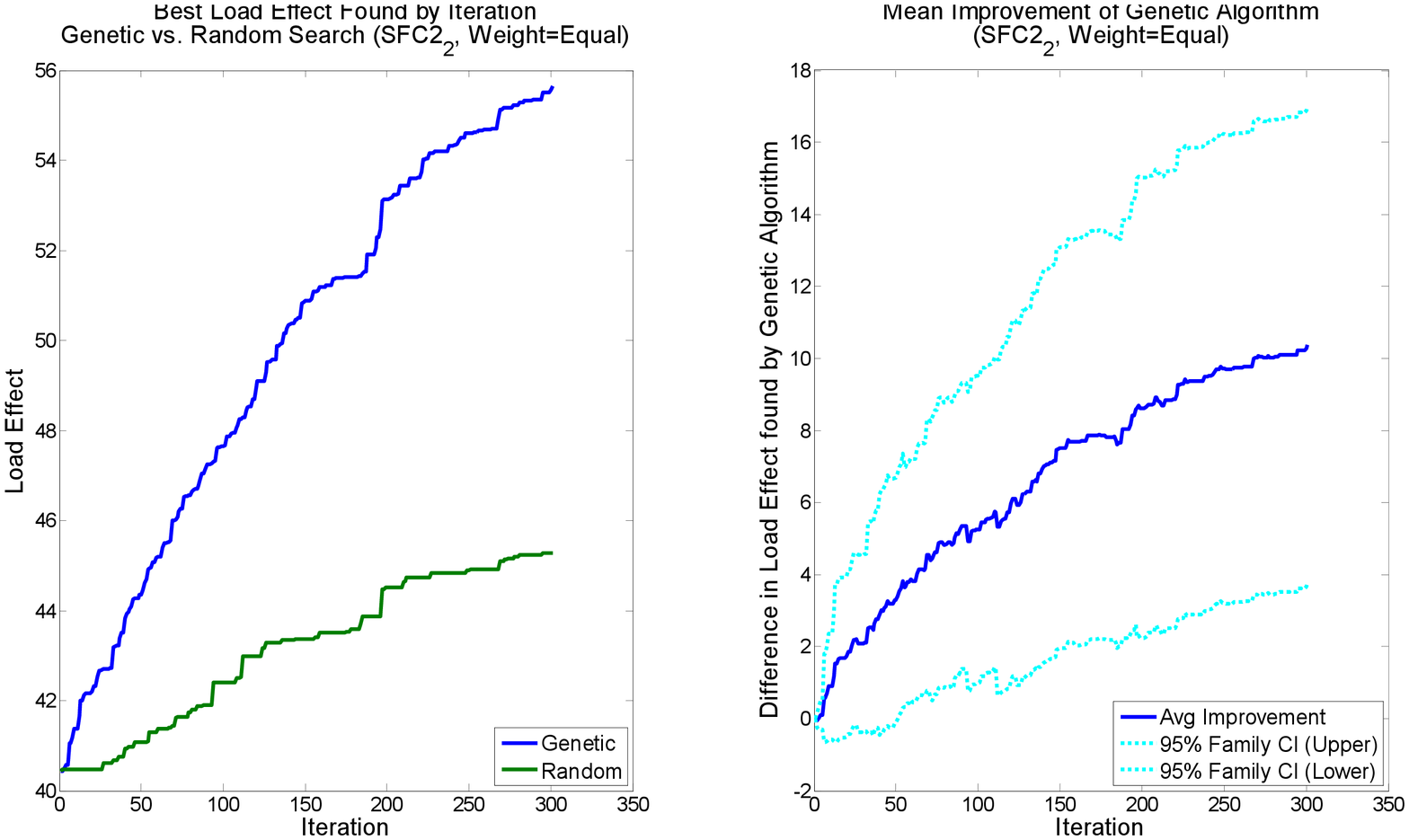}
	\caption{Performance of the genetic algorithm as compared to random search on 276 instances Holme-Kim power law with clustering graphs. \label{fig:SFC2_2GenAlg}}
	\end{figure}

We compared the performance of our genetic algorithm to a simple random enumeration strategy.  Both were run for 300 iterations over our testbed of 276 100-vertex graphs of each type. They were both initialized with a same solution pool comprised of combinations of the eight vertices in each graph having the highest individual load effects on the key vertex\footnote{The number 8 was chosen because ${8\choose 6} \geq 20$, the size of our initial solution pool.}.  Figures \ref{fig:R1GenAlg}, \ref{fig:SW1GenAlg}, \ref{fig:SF2_3GenAlg} and \ref{fig:SFC2_2GenAlg} show that the genetic algorithm outperformed random generation of solutions on every graph type.  The left panel of each figure shows the best load effect found by the two methods as a function of iteration.  The right panel of each figure shows the difference in performance by iteration, with 95\% family confidence intervals that bound the Type I error over all 300 iterations using the Bonferroni adjustment.  Not only does the genetic algorithm outperform random search by a statistically significant amount at nearly every iteration, but it typically achieves the objective function value found by the $300^{th}$ iteration of random search in fewer than 100 iterations.

\section{Applications} \label{sec:apps}
There are several potential applications for this approach, including covert network disruption and the disruption of telecommunication and power networks.  We use counterterrorism as an example in this section.

Prior to his death, Osama bin Laden was, and now Ayman al-Zawahiri is, considered to be vital to the strategic planning, fundraising and morale of al Qaeda. According to Bruce Hoffman, ``Only by destroying the organization's leadership and disrupting the continued resonance of its radical message can the United States and its allies defeat al Qaeda'' \cite{Hof}.

Unfortunately, key leaders of terrorist organizations typically remain clandestine and are difficult to locate with intelligence. A possible remedy is to attempt to increase the visibility of key operatives by removing more easily accessible members from the network. This could directly reduce terrorist activity, an obviously desirable outcome, and it could force the leader to take a more active role in the planning of future attacks. This latter case should create more opportunities to locate these leaders using intelligence since planning attacks requires that more communication involve the leader, which in turn makes him potentially more visible to counter-terrorism officials.

Operations research methodology has already been brought to bear on problems of disrupting terrorist networks.  Marc Sageman has encouraged the use of mathematical models to understand terrorist networks and has partnered with operations researchers in this endeavor \cite{AtS, Bas, CLK, Kre, QXHMC, R2001,RQZLSWC, Sag:2004}.  Sageman \cite{Sag:2004} was among the first to argue that understanding the social network structure of terrorist organizations is critical to identifying the key leaders as well as countering terrorism. He argues terrorist networks evolve according to preferential attachment, where new jihadists become affiliated with the network via well-connected religious clerics or other community leaders.  Typically, new recruits are socially isolated, ex-patriates from their home country who attach themselves to the jihad network in small cliques of friends \cite{Sag:2003, Sag:2004}.  A few different terrorist networks have been considered in the literature \cite{Bas, Kre:2001, Kre, QXHMC, RQZLSWC, Sag:2004}.
In these networks, members previously understood by intelligence officials to be important figures also have high centrality values\footnote{It is interesting to note, however, that when Osama bin Laden was captured, he was discovered in an isolated compound having only one communication channel with the outside world, through his courier.  In this case, he was functioning as a leaf in the network, and no vertex removal could have increased communication flowing through him.}.  Moreover, analysis of the Global Salafi Jihad terrorist network, which includes al Qaeda, suggests that it exhibits an exponentially-truncated power-law degree distribution with a relatively large clustering coefficient \cite{QXHMC}, such as might be generated by the Holme-Kim model described above.

If we assume the volume of communication through a vertex is a proxy for that corresponding member's visibility to intelligence officers, and communication between pairs of members in the organization is proportional to available paths, then we can apply our framework to identify members of the organization to target so that communication will be diverted through an important but clandestine leader.

\section{Future work} \label{sec:future}

This new framework for network flow diversion and disruption opens up a rich area of future research.

\noindent \textbf{What characterizes good vertices to delete?}\label{subsec:vertexchoice}

Our research to date has explored with limited success a few simple heuristics for vertex removal based on centrality metrics and structural equivalence.   As we have seen, the four graph classes studied exhibited large variability in their amenability to load diversion.  Thus, we continue to seek properties that successfully leverage the specific structure of the network to identify the highest load effect vertices in the graph.  Further research into this area would be beneficial for three reasons. First, understanding the mathematical role certain vertices play in the flow capacity of a graph might yield sociological insights. This has happened with other centrality measures, such as betweenness, which has been found to be linked to factors such as job performance, satisfaction and perceived influence in an organization \cite{Bav, Fre, JoP}. Second, understanding the structural relationship between load effect and individual vertices (Single-LOMAX) might yield insights that would help us identify good subsets to remove (LOMAX).  Third, progress in understanding the interplay between removing vertices from a graph and a graph's flow capacity could offer insight to other applications such as the design of robust telecommunication networks.

\noindent \textbf{Improving the genetic algorithm}

Although our results suggest that the genetic algorithm is a promising approach for solving LOMAX, there are many additional features that are often incorporated into genetic algorithms that we have not included here:
 \begin{itemize}
 \item Generation of ``mutations'':  Mutations are randomly perturbed solutions in the solution pool that are introduced to create diversity.
 \item Strategic recombination: Currently we recombine two solutions in an arbitrary fashion, and we test only four possible recombinations for non-repeating vertices before giving up.  There might be more clever ways to recombine solutions that would improve the performance of the algorithm.
 \item Approximate load computation: Computing the load effect of a solution is the bottleneck operation of our genetic algorithm. As a remedy, one could try using approximate centrality metrics to rapidly estimate the load effect of a targeted subset. If such a heuristic is incorporated into our genetic algorithm, then we could evaluate significantly more solutions without increasing the algorithm's running time and earmark the most promising ones to have their load effect computed using exact techniques.  \cite{CKS} survey algorithmic techniques that could be used to speed up computations of common centrality measures in large social networks. Some of the algorithmic ideas, such as using centrality metrics on small subgraphs of the network as an approximate centrality metric on the original graph could be applicable to network flow centrality.  We have found a correlation between load and betweenness, closeness and degree centrality; the strength of this correlation varies by graph type.  Degree is particularly easy to compute and might serve well within a genetic algorithm as an initial screening metric for choosing genetic traits to propagate.
 \end{itemize}

 Additionally, while the genetic algorithm has proven itself to be substantially more effective than random search, other metaheuristics might be worth investigation.

\noindent \textbf{Extensions}

There are also several extensions that could spur a long line of future work in this area.
\begin{itemize}
\item Optimization: In this paper, we have focused on identifying vertices having high load effect on the key vertex without considering whether they are easy targets to remove from the graph.  An optimization version of this problem would include a cost function representing the difficulty of removing a vertex or subset of vertices, and a budget constraint restricting the choice of subsets.
\item Cascading failures: The disruption technique described in this paper focuses on the network at one snapshot in time and assumes that any subset removals occur simultaneously.  However, another modeling approach would be to assume that vertex removals occur sequentially, and we would like to determine the best sequence of vertices to remove to consistently increase the load on the key vertex, similar to the literature on \textit{cascading failures} \cite{CLM, MoL,ZPLY}.
\item Game theory and dynamic response: In a similar vein, we might assume that the terrorist network is continually evolving and can respond strategically to network disruptions by creating new links in the network.  This then leads us to network game theory approaches to this problem.
\item Imperfect information: We assume complete and perfect knowledge of the network's structure.  However there are several extensions to this work that could accommodate imperfect information.  For instance, rather than using unit edge weights, we can assume fractional edge weights that represent the likelihood that an edge exists.
\item Robust network design: We can use the results of this research to design networks, such as power networks, to be robust to load diverting attacks.
\end{itemize}

Additionally, this work could be extended to networks having directed edges or general edge capacities, or networks in which some pairs of vertices carry greater weight in the load calculation than others.  These generalizations could serve to extend the applicability of the work to new contexts.

\section{Conclusions} \label{sec:conc}
In this paper we have presented a new framework for network disruption that is not based on graph connectivity or shortest path lengths but on the amount of flow passing through a key vertex in the network.  For Single-LOMAX, the problem of identifying the single vertex having maximum load effect on the key vertex, we have presented brute force results indicating the potential benefits of this approach and theoretical properties that can help to weed out unpromising targets.  For LOMAX, the problem of identifying a subset of vertices having maximum load effect on the key vertex, we have presented a genetic algorithm that significantly outperforms random search.  We have also listed several extensions that highlight the potentially rich avenues of future research that can be pursued in this area.

This research has broad applicability to problems including disrupting organized crime rings, such as those in terrorism, drug smuggling and human trafficking; disrupting telecommunications networks and power networks; as well as robust network design. Furthermore, this work bridges research in network diversion and social networks; something that has not yet been done.

%

\bibliographystyle{plainnat}
\bibliography{SocialNetsv7}

\section{Appendix}
In this appendix we restate and prove the theorems presented in Section \ref{subsec:proofs}.

\noindent \textbf{Theorem \ref{theorem:kDegree2}.} \textit{Given a graph $G$ with key vertex $k$ having degree $2$, and vertex $i$ adjacent to $k$, $\mathcal{E}_k(G, \{i\}) \leq 0$.}

\begin{proof}
By removing vertex $i$ adjacent to $k$, $k$ becomes a leaf and has load $0$.  Since load is always non-negative, a load on $k$ of $0$ after removing $i$ can be no greater than the load on $k$ in the original graph. \renewcommand{\qedsymbol}{$\blacksquare$}
\end{proof}

\noindent \textbf{Theorem~\ref{theorem:nCycleLoad}.} \textit{Let $G$ be a simple $n$-cycle.  Then for any choice of key vertex $k$ and vertex $i \neq k$ in $G$, $\mathcal{E}_k(G, \{i\}) \leq 0$.}

\begin{proof}[Proof]
We start with the case $n = 3$.  In this case, the load on $k$ in the original graph equals one (the other pair of vertices excluding $k$ has two edge disjoint communication paths, one of which must pass through $k$). Upon removing vertex $i$ from the graph, $k$ becomes a leaf and its load decreases to zero.

For $n \geq 4$, we start by calculating the load on $k$ in the original graph.  Each of the ${n-1 \choose 2}$ origin/destination pairs that exclude $k$ has two edge-disjoint communication paths, one going in either direction around the cycle.  Exactly one of these must include $k$, so the load on $k$ in $G$ is $\mathcal{L}_k(G) = {n-1 \choose 2}$.

Next, we calculate the load on $k$ in the graph with vertex $i$ removed.  There are now ${n-2 \choose 2}$ origin/destination pairs that exclude $k$, and each of these has only a single communication path because the graph $G \setminus \{i\}$ is a tree.  Depending on the proximity of $i$ to $k$ in the original cycle, some of these pairs will communicate through $k$ and some will not.  However, we know that $\mathcal{L}_k(G \setminus \{i\})$ is at most ${n-2 \choose 2}$, which is smaller than ${n -1 \choose 2}$ for $n \geq 4$.
\renewcommand{\qedsymbol}{$\blacksquare$}
\end{proof}

Although the following theorem is a special case of Theorem \ref{theorem:nEDpaths}, we present its proof as it provides an exact calculation of the load effect on $k$ of removing vertex $i$.

\noindent  \textbf{Theorem \ref{theorem:onepath}.} \textit{Let $k$ and $i$ be distinct vertices in graph $G$.  If there is only one edge-disjoint path between $k$ and $i$, then $\mathcal{E}_k(G, \{i\}) \leq 0$.}

\begin{proof}[Proof]
Because $k$ and $i$ have only one edge-disjoint path between them, there exists at least one edge in $G$ whose removal partitions $G$ into two disconnected components, one containing $k$ and one containing $i$.  Let $G_k$ be the subgraph of $G$ over the set of vertices in the component containing $k$, and let $G_i$ be the subgraph of $G$ over the set of vertices in the component containing $i$.  We will show that the load effect on $k$ of removing $i$, $\mathcal{E}_k(G, \{i\})$, is non-positive.

We start by calculating the load on $k$ in the original graph: $\mathcal{L}_k(G) = Z_k(G)-Z_k(G \setminus \{k\})$.  Let $F_{k,G}(G_k)$ denote the flow capacity in $G$ with respect to $k$ over pairs of vertices in $G_k$, $F_{k,G}(G_i)$ denote the flow capacity in $G$ with respect to $k$ over pairs of vertices in $G_i$, and let $F_{k,G}(G_i, G_k)$ denote the flow capacity in $G$ with respect to $k$ between pairs of vertices such that one vertex is in $G_i$ and one vertex is in $G_k$. The flow capacity in $G$ with respect to $k$ is therefore $Z_k(G) = F_{k,G}(G_i)+F_{k,G}(G_k)+F_{k,G}(G_i,G_k)$.  Because only one edge joins $G_k$ and $G_i$ in $G$, communication between pairs of vertices in $G_i$ cannot involve vertices in $G_k$, and vice versa; thus $F_{k,G}(G_i)=Z_k(G_i)$ and $F_{k,G}(G_k)=Z_k(G_k)$.  Moreover, $F_{k,G}(G_i,G_k)=|G_k \setminus \{k\}||G_i|$.    So $Z_k(G) = Z_k(G_i)+Z_k(G_k) + |G_k \setminus \{k\}||G_i|$.

$Z_k(G \setminus \{k\})$, the flow capacity of $G\setminus \{k\}$, is calculated similarly.  Let $s$ denote the number of vertices in $G_k$ that become disconnected from $G_i$ when $k$ is removed from $G$ (excluding $k$).  Thus $Z_k(G\setminus \{k\}) = Z_k(G_k\setminus \{k\}) + Z_k(G_i) + (|G_k \setminus \{k\}| - s)|G_i|.$  The load of $k$ in $G$ is therefore $\mathcal{L}_{G}(k) = Z_k(G)-Z_k(G \setminus \{k\}) = Z_k(G_k) - Z_k(G_k\setminus \{k\}) + s|G_i|$.

We must now calculate the load of k in $G \setminus \{i\}$.  We start by calculating the flow capacity in $G \setminus \{i\}$ with respect to $k$.  Let $p$ denote the number of vertices in $G_i$ that become disconnected from $G_k$ when $i$ is removed (including $i$).  Thus $Z_k(G\setminus \{i\}) = Z_k(G_k) + Z_k(G_i\setminus \{i\}) + |G_k \setminus \{k\}|(|G_i| - p)$  Similarly, when we subsequently remove $k$, $Z_k(G\setminus \{i,k\}) = Z_k(G_k\setminus \{k\}) + Z_k(G_i\setminus \{i\}) + (|G_k \setminus \{k\}| - s)(|G_i| - p).$

The load on $k$ when $i$ is removed is therefore $\mathcal{L}_k(G\setminus \{i\}) = Z_k(G \setminus \{i\})-Z_k(G \setminus \{i,k\}) = Z_k(G_k) - Z_k(G_k\setminus \{k\}) + s(|G_i| - p)$.  The load effect on $k$ of removing vertex $i$ is
\begin{eqnarray}
\mathcal{E}_k(G,i) = \mathcal{L}_k(G\setminus \{i\}) - \mathcal{L}_k(G) =-sp \leq 0
\end{eqnarray}  Thus, the removal of any vertex with exactly one edge-disjoint path to a chosen key vertex $k$ cannot increase the load of $k$.
\renewcommand{\qedsymbol}{$\blacksquare$}
\end{proof}

\noindent \textbf{Theorem \ref{theorem:nEDpaths}.} \textit{Let $k$ and $i$ be distinct vertices in graph $G$.  Consider an edge cut $C$ that partitions $G$ into two components such that $i$ and $k$ are in separate components.  Let $G_k$ be the subgraph of $G$ over the set of vertices in the component containing $k$, and let $G_i$ be the subgraph of $G$ over the set of vertices in the component containing $i$.  Let $i_1,...,i_p$ be the vertices on the $i$ side of the cut that are adjacent to $G_k$.  Let $k_1,...,k_s$ be the vertices on the $k$ side of the cut adjacent to $G_i$.  Suppose any boundary vertex $i_1 \in G_i$ has at least $\lfloor |C|/2 \rfloor$ edge-disjoint paths to every other boundary vertex of $G_i$ by using only vertices in $G_i\setminus \{i\}$, and any boundary vertex $k_1 \in G_k$ has at least $\lfloor |C|/2 \rfloor$ edge-disjoint paths to every other boundary vertex of $G_k$ by using only vertices in $G_k\setminus \{k\}$.  Then $\mathcal{E}_k(G, \{i\}) \leq 0$.}

\begin{proof}[Proof]

First, we note that the condition of any single boundary vertex $i_1$ (respectively $k_1$) having at least $\lfloor |C|/2 \rfloor$ edge-disjoint paths to every other boundary vertex of $G_i$ (respectively $G_k$) by using only vertices in $G_i\setminus \{i\}$ (respectively $G_k \setminus \{k\}$) implies \textit{every pair} of boundary vertices $(i_m,i_n)$ (respectively $(k_m,k_n)$) is connected by at least $\lfloor |C|/2 \rfloor$ edge-disjoint paths using only vertices in $G_i\setminus \{i\}$ (respectively $G_k\setminus \{k\}$).  This is due to a result by Gomory and Hu \cite{GH1961} that the maximum flow between two vertices $x$ and $z$ is at least as large as the minimum of the maximum flow between $x$ and some vertex $y$, and the maximum flow between $y$ and $z$.  The flow between any two boundary vertices $i_m$, and $i_n$ is at least as large as the minimum of the flow between $i_m$ and $i_1$ (which is at least $\lfloor |C|/2 \rfloor$) and the flow between $i_1$ and $i_n$ (which is also at least $\lfloor |C|/2 \rfloor$).  We will use this fact throughout the proof.

The load effect on $k$ of removing vertex $i$ is $\mathcal{E}_k(G, \{i\}) = [ Z_k(G \setminus \{i\}) - Z_k(G \setminus \{i,k\}) ] - [Z_k(G) - Z_k(G \setminus \{k\}) ].$  As in the proof of Theorem \ref{theorem:onepath}, let $F_{k,G}(G_k)$ denote the flow capacity in $G$ with respect to $k$ over pairs of vertices in $G_k$, $F_{k,G}(G_i)$ denote the flow capacity in $G$ with respect to $k$ over pairs of vertices in $G_i$, and let $F_{k,G}(G_i, G_k)$ denote the flow capacity in $G$ with respect to $k$ between pairs of vertices such that one vertex is in $G_i$ and one vertex is in $G_k$.  Note that unlike the proof of Theorem \ref{theorem:onepath}, $F_{k,G}(G_k)$ and $F_{k,G}(G_i)$ are not equal to $Z_k(G_k)$ and $Z_k(G_i)$.  This is because communication between pairs of vertices in $G_k$ can now involve vertices in $G_i$ and vice versa, using edges across the cut $C$.    Then $Z_k(G) =  F_{k,G}(G_k)+ F_{k,G}(G_i) + F_{k,G}(G_i, G_k).$

First consider $F_{k,G}(G_k)$. The number of edge-disjoint paths in $G$ between any two vertices in $G_k$ (neither of which are $k$) that could pass through $G_i$ is $\lfloor |C|/2 \rfloor$, as there are exactly $|C|$ edges in our cut, and each path must enter and leave $G_i$ on a different edge.  Because there are at least $\lfloor |C|/2 \rfloor$ edge-disjoint paths between any pair $(i_m, i_n)$ on the boundary of that cut using only vertices in $G_i \setminus \{i\}$, all flow between any two vertices in $G_k$ that must pass through $G_i$ in $G$ can still pass through $G_i$ in $G \setminus \{i\}$.  Thus, we can conclude that $F_{k, G}(G_k) = F_{k, G \setminus \{i\}}(G_k).$

Next consider $F_{k, G \setminus \{k\}}(G_k)$. By the same logic as above, all flow between two vertices in $G_k \setminus \{k\}$ that must pass through $G_i$ in $G \setminus \{k\}$ can still pass through $G_i$ in $G \setminus \{i,k\}$. Thus we can say that
$F_{k, G \setminus \{k\}}(G_k) = F_{G \setminus \{k,i\}}(G_k).$  The load effect on $k$ of removing $i$, with respect only to pairs of vertices within $G_k$, is therefore zero.

We now consider $F_{k, G}(G_i)$.  By the same logic as above, $F_{k,G}(G_i) = F_{k, G \setminus \{k\}}(G_i)$, and $F_{k,G \setminus \{i\}}(G_i) = F_{k, G \setminus \{k,i\}}(G_i)$.  As a consequence, the load effect on $k$ of removing $i$, with respect only to pairs of vertices within $G_i$, is also zero.

The load effect on $k$ of removing vertex $i$ is now reduced to considering only flows between $G_i$ and $G_k$ across the cut: $$\mathcal{E}_k(i) = [ F_{k,G \setminus \{i\}}(G_i, G_k) - F_{k,G \setminus \{i, k\}}(G_i, G_k)] - [F_{k,G}(G_i, G_k) - F_{k,G\setminus \{k\}}(G_i, G_k) ].$$  We will show this to be non-positive, which is equivalent to showing $$F_{k,G}(G_i, G_k) - F_{k, G \setminus \{i,k\}}(G_i, G_k) \leq [ F_{k,G}(G_i, G_k) - F_{k,G \setminus \{i\}}(G_i, G_k)] + [F_{k,G}(G_i, G_k) - F_{k,G\setminus \{k\}}(G_i, G_k) ].$$  That is, we must show that no additional paths are eliminated when both $i$ and $k$ are removed than are eliminated when $i$ or $k$ are individually removed.

  Because each $F$ term is a summation over all pairs of vertices $a$ and $b$ such that $a \in G_i \setminus \{i\}$ and $b\in G_k\setminus \{k\}$, we will show that the inequality above holds for every such pair $a$ and $b$ via the contrapositive.

  Suppose the theorem conditions hold and there exists an $a-b$ pair such that $$F_{k,G}(a,b)-F_{k,G \setminus \{i\}, k}(a,b)>F_{k,G}(a,b)-F_{k,G \setminus \{i\}}(a,b) + F_{k,G}(a,b)-F_{k,G\setminus \{k\}}(a,b).$$  In words, at least one path from $a$ to $b$ is lost when both $i$ and $k$ are removed despite having remained in a (possibly) rerouted form when either $i$ or $k$ was removed individually.  This is the situation shown in Figure \ref{fig:ProofFig1}.  Let $P$ be the original path in $G$, shown in blue in Figure \ref{fig:ProofFig1_G} and let $P_k$ and $P_i$ be the rerouted versions of $P$ in $G \setminus \{k\}$ and $G \setminus \{i\}$, respectively.  $P$ is identical to exactly one of $P_i$ and $P_k$, and suppose without loss of generality that $P = P_i$.  $P_k$ is the green dotted path in Figure \ref{fig:ProofFig1_G_k}, and $P_i$ is the blue path in both Figures \ref{fig:ProofFig1_G} and \ref{fig:ProofFig1_G_i}.  Then there is a portion of $P$ (and $P_i$) on the $G_i$ side of the cut that avoids vertex $i$.  Likewise, there is a portion of $P_k$ on the $G_k$ side of the cut that avoids vertex $k$.  Moreover, $P_i$ and $P_k$ must overlap on some edges otherwise they could both have been used simultaneously in $G$ to achieve a higher maximum flow.    Because $P_G$ is lost when both $i$ and $k$ are removed despite having been rerouted when either $i$ or $k$ was removed individually, there is no way to match up the portion of $P_i$ on the $G_i$ side of the cut with the portion of $P_k$ on the $G_k$ side of the cut to create a path that avoids both $i$ and $k$ in $G \setminus \{i,k\}$.  This is shown in Figure \ref{fig:ProofFig1_G_ik}.  $P_i$ and $P_k$ must therefore traverse the cut using different edges, and so the maximum flow between $a$ and $b$ in $G$ is at most $|C|-1$.  Since both $P_i$ and $P_k$ are lost when both $i$ and $k$ are removed, the maximum flow in $G \setminus \{i,k\}$ cannot exceed $|C|-2$.

   Suppose in the hardest case that the maximum flow between $a$ and $b$ in $G$ equals $|C|-1$, and the maximum flow in $G \setminus \{i,k\}$ equals $|C|-2$.  Consider the $|C|-2$ paths in $G$ that exclude $P$ (which is the same as $P_i$) and $P_k$.  Each of these paths travels from $a$ to a boundary vertex in $G_i$, across the cut along a distinct edge, to another boundary vertex in $G_k$ and on to $b$.  Any pair of these paths in $G_i$ can be stitched together to create an edge-disjoint path from one boundary vertex to another, passing through $a$ and avoiding $i$.  Likewise any pair of these paths in $G_k$ can be stitched together to create an edge-disjoint path from one boundary vertex to another, passing through $b$ and avoiding $k$.  If $|C|$ is even, then at most $\lfloor |C|/2 \rfloor - 1$ edge-disjoint paths from one boundary vertex to another are used.  The fact that we are unable to stitch together $P_i$ on $G_i$ with $P_k$ on $G_k$ means that there must be strictly fewer than $\lfloor |C|/2 \rfloor$ edge-disjoint paths between the boundary vertices of $P_i$ and $P_k$, and the theorem is proven.  If $|C|$ is odd, then we can pair only $|C|-3$ paths together, creating at most $(|C|-1)/2 -1 = \lfloor |C|/2 \rfloor -1$ edge-disjoint paths from one boundary vertex to another.  However, because the maximum flow in $G \setminus \{i,k\}$ is $|C|-2$ and not $|C|-1$, $a$ and $b$ do not lie on a same cycle on the residual graph consisting of the unpaired path, the portion of $P_i$ on $G_i$ and the portion of $P_k$ on $G_k$ and any unused edges remaining after the first $|C|-3$ units of flow have been pushed.  This is seen in Figure \ref{fig:ProofFig1_G_ik}. If such a cycle existed, it would consist of two edge-disjoint paths from $a$ to one or two boundary vertices of $G_i$, two distinct edges across the cut connecting those boundary vertices to one or two boundary vertices in $G_k$ and two edge-disjoint paths from these boundary vertices of $G_k$ to $b$.  Since this cannot exist, at least one pair of boundary vertices must have strictly fewer than $\lfloor|C|/2\rfloor$ edge-disjoint paths, which we can see in Figure \ref{fig:ProofFig1_G}.  We have proven the contrapositive.

   Therefore, $i$ cannot have a positive load effect on $k$.
 \renewcommand{\qedsymbol}{$\blacksquare$}
\end{proof}

\end{document}